\documentclass[aps,prb,reprint,superscriptaddress,floatfix]{revtex4-2}

\usepackage{amssymb}
\usepackage{graphicx}   
\usepackage{color}      
\usepackage{hyperref}   

\def\lsco{La$_{2-x}$Sr$_x$CuO$_4$}

\def\lbcoate{La$_{1.875}$Ba$_{0.125}$CuO$_4$}

\def\lco{La$_2$CuO$_4$}

\def\newr{\color{black}}

\begin{document}

\title{Spin canting and lattice symmetry in La$_2$CuO$_4$}

\author{Xiao Hu}
\affiliation{Condensed Matter Physics and Materials Science Division, Brookhaven National Laboratory, Upton, New York 11973-5000, USA}
\author{A. Sapkota}
\affiliation{Ames Laboratory, Iowa State University, Ames, IA 50011, USA.}
\affiliation{Condensed Matter Physics and Materials Science Division, Brookhaven National Laboratory, Upton, New York 11973-5000, USA}
\author{V. O. Garlea}
\affiliation{Neutron Scattering Division, Oak Ridge National Laboratory, Oak Ridge, Tennessee 37831, USA}
\author{G. D. Gu}
\author{I. A. Zaliznyak}
\affiliation{Condensed Matter Physics and Materials Science Division, Brookhaven National Laboratory, Upton, New York 11973-5000, USA}
\author{J. M. Tranquada}
\email{jtran@bnl.gov}
\affiliation{Condensed Matter Physics and Materials Science Division, Brookhaven National Laboratory, Upton, New York 11973-5000, USA}

\date{\today} 

\begin{abstract}
While the dominant magnetic interaction in La$_2$CuO$_4$ is superexchange between nearest-neighbor Cu moments, the pinning of the spin direction depends on weak anisotropic effects associated with spin-orbit coupling.  The symmetry of the octahedral tilt pattern allows an out-of-plane canting of the Cu spins, which is compensated by an opposite canting in nearest-neighbor layers.  A strong magnetic field applied perpendicular to the planes can alter the spin canting pattern to induce a weak ferromagnetic phase.  In light of recent evidence that the lattice symmetry is lower than originally assumed, we take a new look at the nature of the field-induced spin-rotation transition.  Comparing low-temperature neutron diffraction intensities for several magnetic Bragg peaks measured in fields of 0 and 14 T, we find that a better fit is provided by a model in which spins {\newr rotate within both} neighboring {\newr planes but by different amounts}, resulting in a noncollinear configuration.  This model allows a {\newr more} consistent relationship between lattice symmetry and spin orientation at all Cu sites.
\end{abstract}

\maketitle

\section{Introduction}

Interest in \lco, the parent compound of the original high-temperature superconductor family \cite{bedn87,keim15}, was renewed recently with the discovery by Taillefer and coworkers that, at low temperature, it exhibits an unusually large thermal Hall conductivity \cite{gris19}.  The original measurements were performed in a $c$-axis magnetic field of 15 T, which puts the system into the weak ferromagnetic phase, associated with a field-induced in-phase alignment of the small out-of-plane canting of the Cu spins \cite{kast88,reeh06}.  Further studies have provided evidence that the thermal Hall conductivity is due to phonons \cite{gris20} and that a sizable response can also be found in nonmagnetic compounds such as SrTiO$_3$ \cite{li20}.  While a number of possible explanations based on intrinsic effects have been proposed \cite{sama19,han19,varm20}, extrinsic effects involving skew scattering off of defects might be the dominant effect \cite{chen20,guo21}.  Nevertheless, experimentalists have discussed the effect in terms of chiral phonons \cite{gris20,boul20}.

Following on work by Reehuis {\it et al.}\ \cite{reeh06}, we recently demonstrated \cite{sapk21} that the structural symmetry of \lco\ is lower than originally determined \cite{lehu73,long73}.  In particular, the ordered rotation of the CuO$_6$ octahedra includes a small component rotated around the Cu-O bond axis, with associated anomalous soft phonons.  A complete softening of such phonons occurs in the related compound La$_{1.8}$Eu$_{0.2}$CuO$_4$ below 133~K \cite{huck04}.  Similarities between the magnetization of that low-temperature phase and the field-induced weak-ferromagnetic phase of \lco\ led us to wonder whether the soft phonons might couple to the canted spins in an interesting way (despite the fact that measurements on other cuprates indicate that canted moments are not essential for observation of a large thermal Hall conductivity \cite{boul20}).

We have used neutron scattering to study the impact of a 14-T magnetic field applied along the $c$-axis of a crystal of \lco.  We find that the {\newr crystal} structure is quite stable to the field, and no significant change to phonons was detected.  The measurements did, however, provide an opportunity to reconsider the proposed model for the weak ferromagnetic phase.  We propose a new model involving a noncollinear arrangement of spins in neighboring planes and show that it gives a better fit to our measured magnetic peak intensities.

\begin{figure*}[t]
\centering
\includegraphics[width=1.8\columnwidth]{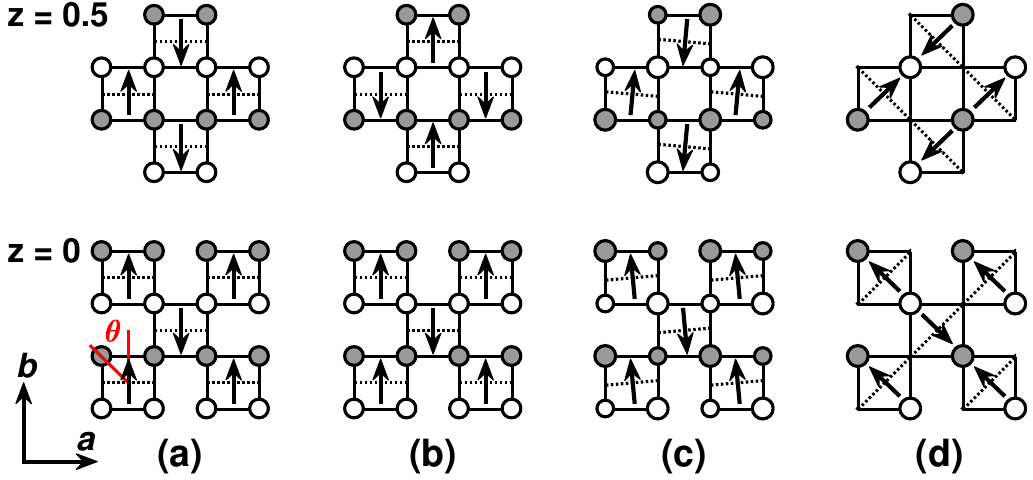}
\caption{\label{fg:spins}\newr Schematics of possible spin configurations in CuO$_2$ layers, where each arrow represents a Cu spin, circles indicate oxygen atoms (white/gray indicates displacement above/below plane; size indicates relative displacement), dashed line indicates octahedral tilt axis, and lower/upper panel corresponds to the plane displaced along the $c$ by $z=0$/$z=0.5$.  The orientations of the $a$ and $b$ axes are indicated at the lower left; the definition of the spin rotation angle $\theta$ is indicated in red.  (a) Conventional model of antiferromagnetic order in zero field.  (b) Conventional model of order in the weak ferromagnetic phase, with spins in the $z=0.5$ layer rotated by 180$^\circ$.  (c) Zero-field model allowing for in-plane canting of spins, following the octahedral tilt direction in the revised structure, plotted here with $\alpha=5^\circ$.  (d) Spin and tilt pattern in low-temperature phase of La$_{1.8}$Eu$_{0.2}$CuO$_4$ exhibiting weak ferromagnetism \protect\cite{huck04}.}
\end{figure*}

To provide context, we note that,
as \lco\ is a charge-transfer correlated insulator \cite{toku90,zaan85}, it exhibits strong antiferromagnetic correlations within the CuO$_2$ planes from high temperatures \cite{shir87,imai93,birg99} driven by a large nearest-neighbor superexchange energy \cite{ande87,cold01}.  Antiferromagnetic order \cite{vakn87,budn87} develops below a N\'eel temperature of $T_{\rm N}\sim325$~K \cite{keim92}, which is sensitive to oxygen stoichiometry \cite{well97}.  While the ordered Cu moments lie largely within the CuO$_2$ planes [see Fig.~\ref{fg:spins}(a)] \cite{vakn87}, there is a small canting perpendicular to the planes that is apparent as a rise in the magnetization (measured with a field perpendicular to the planes) on cooling toward $T_{\rm N}$, where the magnetic layers are decoupled; below $T_{\rm N}$, the canting in neighboring layers cancels out, resulting in a decrease in the magnetization \cite{thio88,lavr01}.  {\newr The approximate magnetic structure and its relation to the octahedral tilt pattern are illustrated in Fig.~\ref{fg:spins}(a).}

The spin canting has been explained \cite{thio88} as a consequence of 
exchange terms resulting from the effects of spin-orbit coupling as originally identified by Dzyaloshinsky \cite{dzya58} and Moriya \cite{mori60} (DM).  The evaluation of the DM interaction for cuprates with particular lattice symmetries is not trivial, and it received considerable attention in the decade following the discovery of cuprate superconductivity \cite{coff91,shek92,bone93,enti94,kosh94,vier94b,yild94b,stei96}.  

Empirically, the moment direction and canting in zero field are found to be orthogonal to the octahedral rotation axis \cite{keim93,huck04}{\newr; this comes from studies of Nd- and Eu-doped \lco, in which the low-temperature phase has a modified octahedral tilt pattern and a noncollinear spin order, as shown in Fig.~\ref{fg:spins}(d).  In fact, the Eu-doped compound exhibits weak ferromagnetism in the noncollinear  phase at low field, with a magnetization very similar to that of the high-field magnetization of \lco\ \cite{huck04}. This makes it tempting to associate the canting of the moments with the octahedral tilts.  There are two problems with such an association: 1) The size of the canted moment estimated from the high-field magnetization is much smaller than one would expect from the measured moments and tilt angles.  2) The relationship between the spin direction and the octahedral tilt direction reverses from one layer to the next in the zero-field phase, as can be seen in Fig.~\ref{fg:spins}(a).  One might try to explain the latter point in terms of the impact of interlayer exchange; however, that energy is quite small and nearly frustrated \cite{yild94a}.  We note that one feature not considered in the previous analyses is the shear distortion of the octahedra associated with the monoclinic symmetry \cite{reeh06,sapk21}.  That distortion has the same orientation for Cu sites in neighboring planes with the same zero-field spin direction, in contrast to the opposite signs of the tilt directions.
} 

{\newr
If the spin structure of \lco\ were collinear, with the spin direction determined by the interlayer coupling \cite{yild94a}, then it would be reasonable that a high magnetic field could flip the spin in every other layer, as originally proposed \cite{thio88} and as shown in Fig.~\ref{fg:spins}(b).  A challenge, however, is that we now know \cite{sapk21} that the octahedral tilts are rotated slightly away from the $b$ axis, as indicated in Fig.~\ref{fg:spins}(c), and we expect a corresponding noncollinear rotation of the spins.  In fact, we will show that a fit to our neutron diffraction intensities in zero field indicates a finite in-plane rotation angle for the spins.  
}

{\newr
The noncollinear structure suggests that anisotropies due to DM interactions plus Coulomb exchange interactions determine the directions of the ordered spins \cite{huck04,kosh94,yild94b}. 
In such a case, it seems more reasonable that a large applied field will cause spins in both layers to rotate in some fashion.  Indeed, we find that such a field-induced spin rotation model gives a better fit to our high-field neutron diffraction data than does the collinear spin-flip model.}

The rest of the paper is organized as follows.  After a description of the experimental methods, the results are presented in Sec.~III and analyzed in Sec.~IV.  We end with a summary and discussion in Sec.~V.

\section{Experimental Methods}

The \lco\ sample studied here is the same 11-g vacuum-annealed crystal used previously in \cite{sapk21}.  Based on the temperature dependent peak in the magnetic susceptibility, $T_N$ is equal to 327~K.  The lattice parameters of 
{\newr \lco\ at low temperature are $a=5.335$~\AA, $b=5.415$~\AA, and $c=13.12$~\AA\ \cite{rada94}.}
We will express the momentum transfer $Q = (H, K, L)$ in reciprocal-lattice units (r.l.u.) given by $(2\pi/a, 2\pi/b, 2\pi/c)$.

The present neutron scattering experiment was performed on the HYSPEC spectrometer \cite{hyspec15} at the Spallation Neutron Source, Oak Ridge National Laboratory.  The crystal was mounted in the 14-T vertical-field magnet, with the $c$ axis parallel to the magnetic field and transverse to the horizontal scattering plane.  The sample orientation with respect to the incident beam could be changed by rotation about the vertical axis.  An incident energy of 27~meV was selected and the Fermi chopper frequency was set at 300~Hz.  For measurements of the magnetic peak intensities, the center of the movable area detector was set to $34^\circ$.  Data were collected over a range of sample angles with steps of 0.25$^\circ$ between measurements.

\begin{figure}[t]
 \centering
    \includegraphics[width=0.95\columnwidth]{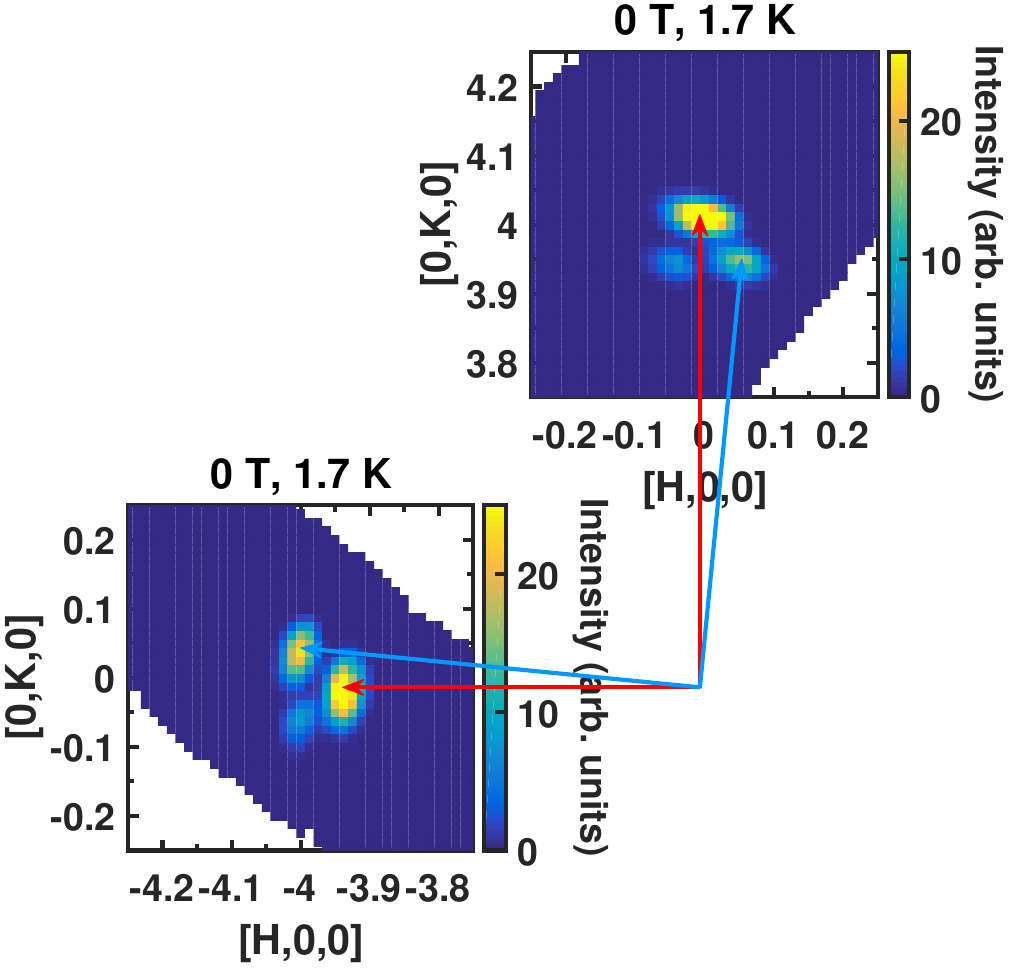}
    \caption{\label{fg:dom}  Maps of intensities around $(\bar{4}00)$ and $(040)$ reflections; red arrows point to the reflections from the dominant domain, while blue arrows correspond to one of the two twin domains. }
\end{figure}

Some measurements at larger {\bf Q} were performed with the detector at $-59^\circ$.  In particular, measurements of the $(040)$ and $(\bar{4}00)$ peaks, with a step size of 0.1$^\circ$, are shown in Fig.~\ref{fg:dom}.  As one can see, there is one dominant domain with two twin domains rotated with respect to it by $\pm89^\circ$.  

For convenience, we will label reflections according to the dominant domain; for the low-$Q$ peaks, the twin domains are not well resolved, so that integrated intensities combine the dominant and twin reflections.  We measured the $(100)$, $(120)$, and $(300)$ reflections, where the dominant domain has magnetic reflections, and $(010)$ and $(210)$ peaks, where the magnetic signal of the twin domains appears.  Data were collected in fields of 0 and 14~T applied along the $c$ axis.

\section{Results}

\begin{figure}[t]
 \centering
    \includegraphics[width=1\columnwidth]{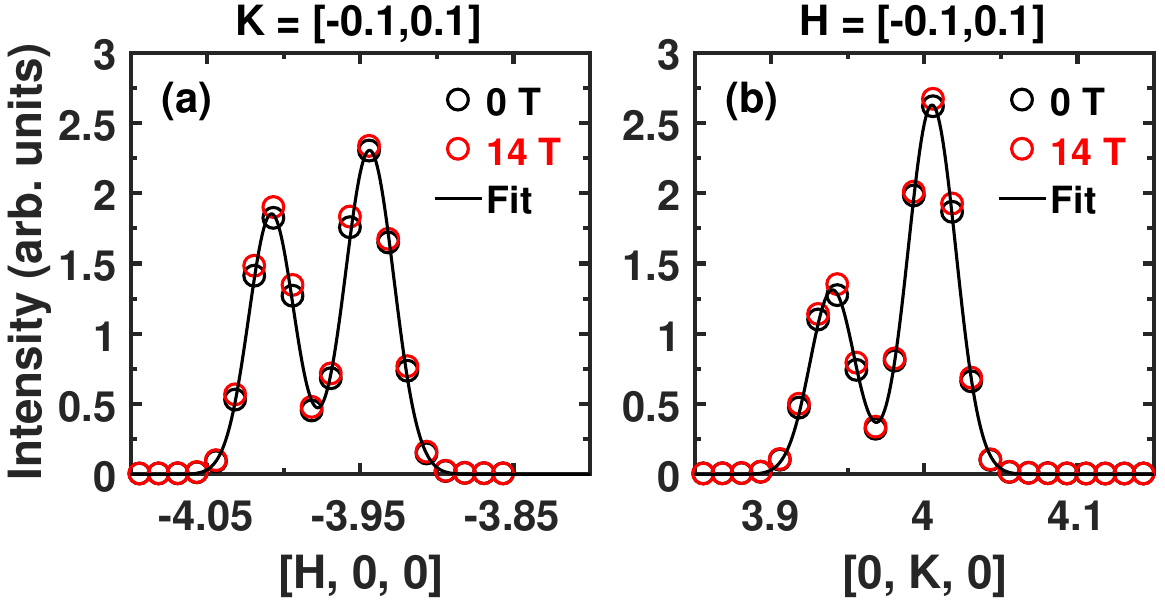}
    \caption{\label{fg:mdep1}  Plot of Bragg peaks near (a) $(-4,0,0)$ and (b) (040), in $B=0$ (black) and $B=14$~T (red).  In (a) [(b)], the data have been integrated over $K=0\pm0.1$ [$H=0\pm0.1$] to catch the intensities from all domains. The absence of any detectable change to the orthorhombicity due to the field is readily apparent.}
\end{figure}

To check the impact of the magnetic field on the crystal structure, we show in Fig.~\ref{fg:mdep1}  nuclear scattering along longitudinal cuts through the $(-4,0,0)$ and $(0,4,0)$ positions (with integration over the transverse directions) in fields of 0 and 14~T.  As one can see, there is no significant change with field of either the orthorhombic splitting or the peak intensities.  

As the structure is unaffected by the field, we turn to the magnetic reflections.  Previous neutron diffraction studies of the field-induced transition to the weak ferromagnetic phase have focused on the (100) and (210) magnetic peaks \cite{kast88,reeh06}.   The integrated intensities of those peaks, along with (010) and (120), and their dependence on field are listed in Table I.  [Note that the (210) and (120) reflections are close to an Al powder ring from the sample holder, so that some careful subtraction of the Al ring had to be done.]  A complication is that there are structural superlattice peaks allowed at (010) and (210) \cite{reeh06,sapk21}, and we were not able to separate the scattering from the twin domains from the main domain.  As a result, modeling of the measured intensities must account for both the magnetic and nuclear scattering from all domains at each of these four peaks.

\begin{table}[b]
	\caption{Integrated intensities of superlattice peaks (in arbitrary units) at $T=1.7$~K vs.\ field. \label{TM}}.
	\begin{ruledtabular} 
		\begin{tabular}{ccc}
			$(HK0)$ & $I(B=0)$ & $I(B=14\,{\rm T})$ \\
			\hline
			(100) & $2.51 \pm 0.07$ & $\ 0.72 \pm 0.07$ \\
			(010) & $3.39 \pm 0.08$ & $\ 3.25 \pm 0.08$ \\
			(120) & $4.53 \pm 0.16$ & $\ 4.65 \pm 0.16$ \\
			(210) & $9.03 \pm 0.19$ & $11.61 \pm 0.20$ \\
			(300) & $1.98 \pm 0.11$ & $\ 0.76 \pm 0.10$ \\
		\end{tabular}
	\end{ruledtabular}
\end{table}

\section{Analysis}

It has been established empirically that the spin direction tends to align with the octahedral tilt direction (perpendicular to the octahedral rotation axis) \cite{keim93,huck04}.   In the low-temperature-orthorhombic (LTO) structure, long believed to characterize \lco, the octahedral tilts point along the $b$ axis.  Neutron diffraction studies have shown that the Cu magnetic moment is largely transverse to the $a$ axis, roughly consistent with this picture, as indicated in Fig.~\ref{fg:spins}(a).  The recent neutron evidence indicates that the octahedral tilt direction deviates slightly from the $b$ axis, with that deviation alternating its orientation from one layer to the next.  

Let us consider two adjacent CuO$_2$ planes.  To describe the spin directions, we consider site $j=1$ in the first layer at ${\bf R} = (0,0,0)$ and site $j=2$ in the second layer at ${\bf R} = (\frac12,0,\frac12)$.  {\newr We will take the spin configuration in Fig.~\ref{fg:spins}(a) as the reference; the spin direction at sites $j=1$ and 2 can then be written as
\begin{eqnarray}
  \hat{\bf S}_1 & = & (-\sin\theta_1,\cos\theta_1,0), \\
  \hat{\bf S}_2 & = & (\sin\theta_2,-\cos\theta_2,0).
\end{eqnarray}
Note that we ignore the $c$-axis component; it was estimated \cite{kast88} to have a canting angle of 0.17$^\circ$, corresponding to $\hat{S}_z = 0.003$, which is too small to impact neutron scattering measurements.  }
The nominal LTO phase {\newr is described by} $\theta_1 = 0$ and {\newr $\theta_2 = 0$}.  We now know that the {\newr octahedral} tilt direction is rotated with respect to the $b$ axis (though the precise rotation angle has not been determined) \cite{sapk21}.  If we call that rotation $\alpha$, then {\newr we expect} $\theta_1 = \alpha$ and $\theta_2 = -\alpha$.   The case of spins following the tilt direction, with {\newr $\alpha=5^\circ$}, is indicated in Fig.~\ref{fg:spins}(c).  

For magnetic diffraction, only the component of spin perpendicular to {\bf Q}{\newr, $\hat{\bf S}_{\perp j}$,} contributes.  
We can then write the magnetic structure factor as
\begin{equation}
F_{\rm mag}(H,K,0) = A f_{\rm Cu}({\bf Q})F_{\rm AF}\left|\hat{\bf S}_{\perp 1} + \hat{\bf S}_{\perp 2}e^{i\pi H}\right|,
\end{equation}
where $A$ is a scale factor (to be determined by fitting) and $F_{\rm AF} = \cos(\pi H) - \cos(\pi K)$.  For the magnetic form factor, $f_{\rm Cu}(\bf Q)$, we use the results of \cite{walt09} to obtain the ${\bf Q}$ dependence \footnote{For the magnetic form factor, we use $f_{\rm Cu}(1,0,0) = f_{\rm Cu}(0,1,0) = 0.63$, $f_{\rm Cu}(2,1,0) = f_{\rm Cu}(1,2,0) = 0.55$, and $f_{\rm Cu}(3,0,0) = 0.50$.}.

There is a nuclear contribution for $(H,K,0)$ with $K$ odd and $H$ even \cite{reeh06}.  From our previous study of the same sample, measurements at 332~K (above $T_N$, where there is no magnetic intensity) we find that $|F_N(2,1,0)|^2/|F_N(0,1,0)|^2 = 3.01$ and $|F_N(0,3,0)|^2/|F_N(0,1,0)|^2 = 0.70$.  We will assume that the ratio remains the same at low temperature, and take $|F_N(0,1,0)|$ as a fitting parameter.

The total structure factor at a given {\bf Q} is then
\begin{equation}
F_{\rm tot} = F_{\rm mag} + F_{\rm N}, 
\end{equation}
and the measured intensities $I$ are proportional to $|F_{\rm tot}|^2$.  Since we integrate over intensities from twin domains, we still need to take account of the inequivalent reflections that are combined in this way. Let the primary domain be associated with $(H,K,0)$; then, we will evaluate the effective intensity as
\begin{equation}
I_{\rm eff}(H,K,0) = I(H,K,0) + c\cdot I(K,H,0),
\end{equation}
where $c$ is the relative intensity of the minority twin domains relative to the main domain for the same {\bf Q}.  The value $c=0.598$ was determined from the intensities of the resolved twin domains for the $(\bar{4}00)$ and (040) reflections.

\begin{figure}[t]
 \centering
    \includegraphics[width=1\columnwidth]{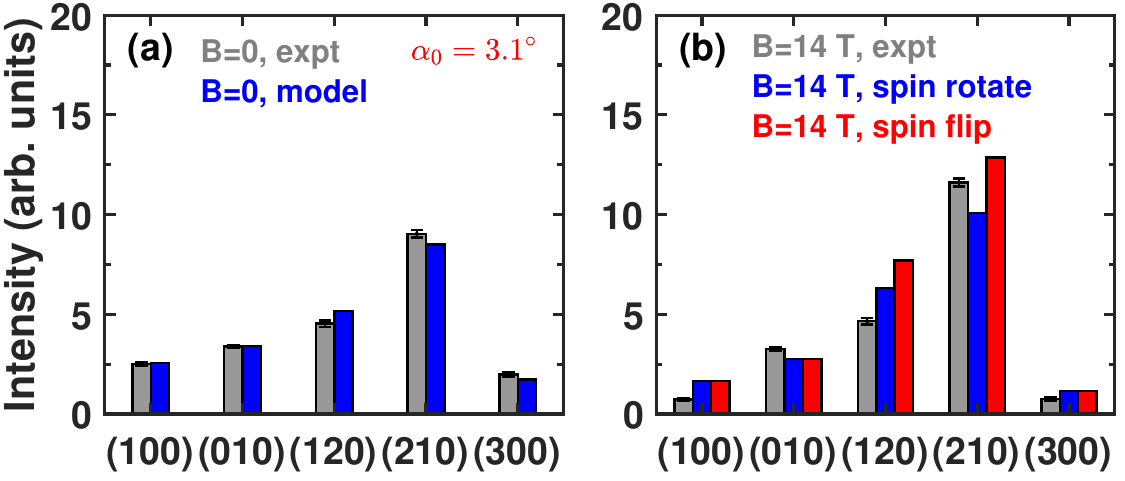}
    \caption{\label{fg:fit}  Comparison of measured superlattice peak intensities (a) at zero field and (b) at 14~T, with fits to the models. }
\end{figure}

{\newr
To check our assumed ground-state spin structure, we fit the spin angle $\theta_0 \equiv \theta_1 = -\theta_2$ along with the unknown parameters, $A$ and $|F_N(0,1,0)|$, to the zero-field data.  We did this by evaluating $\chi^2$ on a 3-dimensional parameter grid with steps of 0.1$^\circ$, 0.001, and 0.01, respectively.  The global minimum, with $\chi^2=15.2$, occurs for $\theta_0 = 3.1^\circ$; for comparison, $\chi^2$ rises to 15.6 at $\theta_0=0$.  If we take an increase of $\chi^2$ by 1 to determine the uncertainty in parameters, then we find that $\theta_0$ could be as large as 8.8$^\circ$.  The fitted intensities are compared with the measurements in Fig.~\ref{fg:fit}(a).  We will take $\theta_0$ as a measure of the octahedral rotation angle, so that $\alpha_0=3.1^\circ$.

It is interesting to note that, in the neutron diffraction study by Reehuis {\it et al.} \cite{reeh06}, it was found that the (010) peak intensity grew on cooling from $T  > T_{\rm N}$ to 10~K.  They noted that if the increase were attributed to magnetic scattering, it would correspond to a moment of 0.060~$\mu_{\rm B}$ along the $a$ axis.  The moment along the $b$ axis was determined to be 0.42~$\mu_{\rm B}$.  In our model, that would correspond to $\theta_0=8.1^\circ$, within the range of uncertainty for our fit.
}

\begin{figure}[t]
 \centering
    \includegraphics[width=0.8\columnwidth]{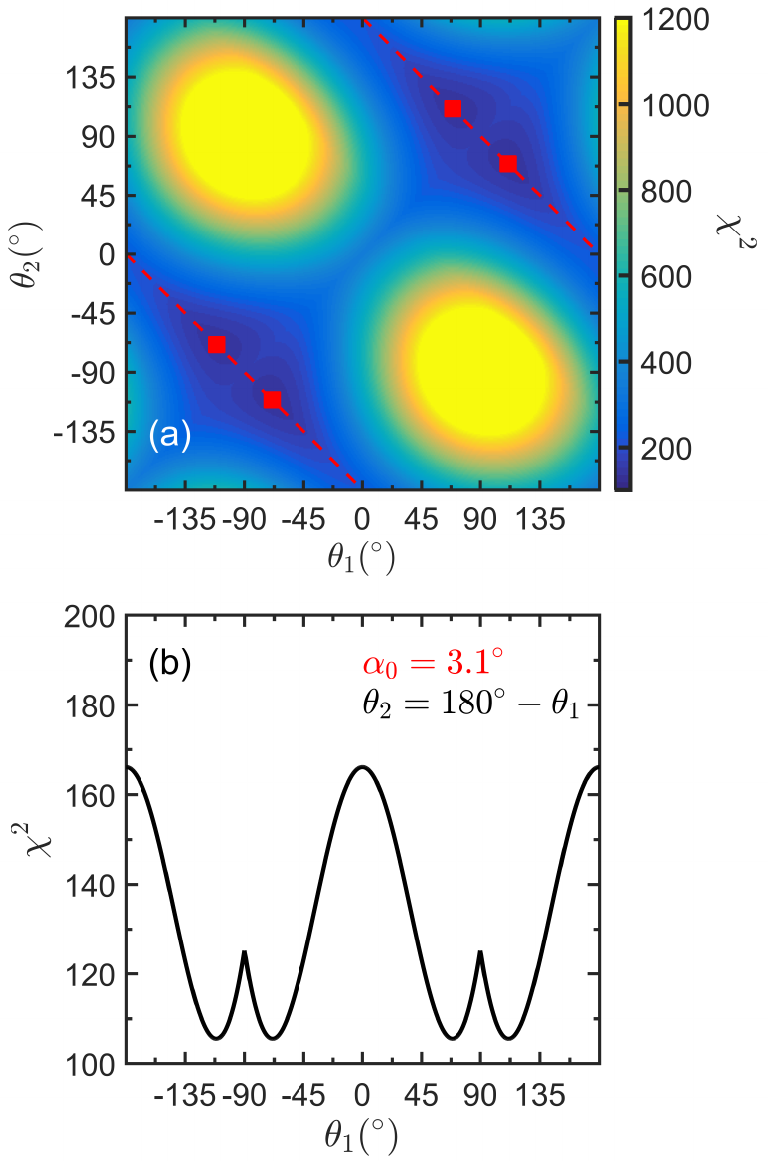}
    \caption{\label{fg:chi} \newr (a) False-color plot of $\chi^2$ vs.\ $\theta_1$ and $\theta_2$ obtained in fitting the high-field superlattice peaks with the noncollinear spin model.  Red dashed lines indicate valleys corresponding to $\theta_2=180^\circ - \theta_1$; red squares indicate absolute minima.  (b) Variation of $\chi^2$ vs.\ $\theta_1$ for the condition $\theta_2 = 180^\circ - \theta_1$. }
\end{figure}

{\newr
Next, we consider the high-field data.  Here we use the values of $A$ and $|F_N(0,1,0)|$  determined from the zero-field data, assuming that there is no significant field dependence.  We initially treated the rotation angles $\theta_1$ and $\theta_2$ as independent parameters.  A plot of $\chi^2$ as a function of these two parameters is shown in Fig.~\ref{fg:chi}(a).  Note that we expect to obtain 4 equivalent minima, as the absolute spin direction is not constrained and exchanging $\theta_1$ and $\theta_2$ does not impact the fit.  We find that the minima occur along the lines $\theta_2=180^\circ -\theta_1$ (modulo 360$^\circ$), indicated by dashed red lines, with the absolute minima indicated by the red squares.  
}

\begin{figure}[t]
 \centering
    \includegraphics[width=0.5\columnwidth]{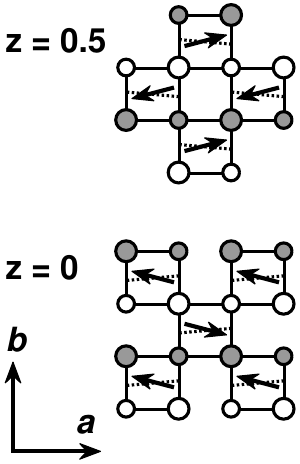}
    \caption{\label{fg:opt}\newr Schematic of the high-field noncollinear spin model with the optimal $\theta_1=69^\circ$, $\theta_2=111^\circ$; symbols have same meaning as in Fig.~\protect\ref{fg:spins}. }
\end{figure}

{\newr
In Fig.~\ref{fg:chi}(b), we plot $\chi^2$ along the line $\theta_2=180^\circ -\theta_1$ as a function of $\theta_1$.  Selecting one of the equivalent minima, we have $\theta_1 = 69^\circ$ and $\theta_2=111^\circ$.  The spin-flip model of Thio {\it et al.} \cite{thio88} corresponds to $\theta_1=0^\circ$, and we see that it gives a much poorer fit to the data.  The calculated peak intensities for the two models are compared with the measured ones in Fig.~\ref{fg:fit}(b).
The optimized high-field spin configuration is shown in Fig.~\ref{fg:opt}.
}




\section{Summary and Discussion}

Previous neutron diffraction tests \cite{kast88,reeh06} of the {\newr magnetic-field-induced} transition to the weak ferromagnetic phase 
observed changes in the intensities of the (100) and (210) magnetic peaks that were qualitatively consistent with the original proposal that Cu spins in half of the CuO$_2$ planes flip their direction {\newr at the transition} \cite{thio88}.   
Our quantitative analysis of several peaks indicates a better fit with a model that treats the layers {\newr in a more} uniform {\newr fashion}, allowing for field-induced spin rotations to result in a noncollinear state.  {\newr We find that the spin orientations in neighboring layers change from $\theta_1+\theta_2=0$ in zero field to $\theta_1+\theta_2=180^\circ$ in high field, where in each case $\theta_1$ and $\theta_2$ are nonzero.  (These equations also describe the spin-flip model, but with $\theta_1=0$ in both field cases.)  }

This result is intriguing on its own, but we think it also has relevance to understanding other experimental observations.  For example, we have already mentioned the peak in the magnetic susceptibility observed at $T_{\rm N}$ when the magnetic field is applied along the $c$ axis \cite{thio88,lavr01}.  That situation became more complicated when measurements on a detwinned crystal revealed that there was also a peak in the susceptibility when the field is applied along the orthorhombic $b$ axis \cite{lavr01}.  A theoretical analysis including both the DM and pseudodipolar direct exchange interactions provided a qualitative description of this effect, assuming that the lattice symmetry corresponds to space group $Bmab$ \cite{silv06}.  We suggest that including the slight in-plane canting of the spin directions due to the small adjustment of the octahedral tilt axes in space group $P2_111$ \cite{sapk21} may lead to better agreement with experiment \cite{lavr01}.

Another related issue is the static spin stripe order observed in \lbcoate\ \cite{fuji04}, which has the low-temperature tetragonal (LTT, space group $P4_2/ncm$) structure \cite{axe89,huck11}.  The pinning of the spin direction generally depends on anisotropy of the exchange interactions, which requires consideration of both spin-orbit and Coulomb exchange interactions \cite{yild94b,kosh94}{\newr; for coupled spin ladders without anisotropic interactions, an unreasonably large exchange coupling between neighboring ladders (across charge stripes) is necessary to achieve order \cite{twor99,yao06a}.}  (A small energy gap of $\sim0.7$~meV in the spin excitations, as expected for anisotropic spin couplings, was observed at low temperature \cite{tran08}.)  For in-plane magnetic field, the anisotropy of the spin-flip transition with field relative to the Cu-O bond directions is consistent with a zero-field orientation of spins perpendicular to the octahedral tilt axes, which corresponds to orthogonal spin directions in neighboring planes \cite{huck08}.  There have been conflicting predictions \cite{kosh94,stei96} for the Cu spin orientation in cuprates with LTT structure, where experimental results \cite{huck04} are more consistent with \cite{kosh94}.  Our new result provides another experimental constraint that requires a consistent understanding.

A related puzzle in \lbcoate\ is the observation of a finite magneto-optical Kerr effect that onsets with the transition to the LTT phase \cite{kara12} where charge-stripe order onsets \cite{fuji04,huck11,wilk11}.  It was measured with a technique such that the finite Kerr effect implies time-reversal symmetry breaking \cite{oren11,frie14}.  Further measurements provided evidence for chirality in association with the Kerr effect \cite{kara14}.  Theoretical analyses have considered the possibility of chiral-nematic charge order \cite{wang14} as well as charge stripes plus spin-orbit coupling \cite{oren13}.  DM interactions are a logical spin-orbit effect to expect, and combining nematic spin-stripe order with the charge stripes might provide a basis for understanding the Kerr effect results, especially if there are interesting correlations between neighboring layers along the $c$ axis.  
{\newr A very recent x-ray diffraction study \cite{sear23} has provided evidence of a structural coupling between charge stripes in neighboring layers.  Whether this coupling, plus spin-orbit effects, can provide the chirality needed to explain the Kerr-effect results will require further theoretical work.
}

\medskip
\section{Acknowledgments}

We thank Melissa Graves-Brook for assistance with the experiment at HYSPEC.
Work at Brookhaven is supported by the Office of Basic Energy Sciences, Materials Sciences and Engineering Division, U.S. Department of Energy (DOE) under Contract No.\ DE-SC0012704.   A portion of this research used resources at the Spallation Neutron Source and the High Flux Isotope Reactor, DOE Office of Science User Facilities operated by Oak Ridge National Laboratory.

\bibliography{LNO,theory,neutrons}

\begin{thebibliography}{62}%
\makeatletter
\providecommand \@ifxundefined [1]{%
 \@ifx{#1\undefined}
}%
\providecommand \@ifnum [1]{%
 \ifnum #1\expandafter \@firstoftwo
 \else \expandafter \@secondoftwo
 \fi
}%
\providecommand \@ifx [1]{%
 \ifx #1\expandafter \@firstoftwo
 \else \expandafter \@secondoftwo
 \fi
}%
\providecommand \natexlab [1]{#1}%
\providecommand \enquote  [1]{``#1''}%
\providecommand \bibnamefont  [1]{#1}%
\providecommand \bibfnamefont [1]{#1}%
\providecommand \citenamefont [1]{#1}%
\providecommand \href@noop [0]{\@secondoftwo}%
\providecommand \href [0]{\begingroup \@sanitize@url \@href}%
\providecommand \@href[1]{\@@startlink{#1}\@@href}%
\providecommand \@@href[1]{\endgroup#1\@@endlink}%
\providecommand \@sanitize@url [0]{\catcode `\\12\catcode `\$12\catcode
  `\&12\catcode `\#12\catcode `\^12\catcode `\_12\catcode `\%12\relax}%
\providecommand \@@startlink[1]{}%
\providecommand \@@endlink[0]{}%
\providecommand \url  [0]{\begingroup\@sanitize@url \@url }%
\providecommand \@url [1]{\endgroup\@href {#1}{\urlprefix }}%
\providecommand \urlprefix  [0]{URL }%
\providecommand \Eprint [0]{\href }%
\providecommand \doibase [0]{https://doi.org/}%
\providecommand \selectlanguage [0]{\@gobble}%
\providecommand \bibinfo  [0]{\@secondoftwo}%
\providecommand \bibfield  [0]{\@secondoftwo}%
\providecommand \translation [1]{[#1]}%
\providecommand \BibitemOpen [0]{}%
\providecommand \bibitemStop [0]{}%
\providecommand \bibitemNoStop [0]{.\EOS\space}%
\providecommand \EOS [0]{\spacefactor3000\relax}%
\providecommand \BibitemShut  [1]{\csname bibitem#1\endcsname}%
\let\auto@bib@innerbib\@empty
\bibitem [{\citenamefont {Bednorz}\ \emph {et~al.}(1987)\citenamefont
  {Bednorz}, \citenamefont {Takashige},\ and\ \citenamefont
  {M\"uller}}]{bedn87}%
  \BibitemOpen
  \bibfield  {author} {\bibinfo {author} {\bibfnamefont {J.~G.}\ \bibnamefont
  {Bednorz}}, \bibinfo {author} {\bibfnamefont {M.}~\bibnamefont {Takashige}},\
  and\ \bibinfo {author} {\bibfnamefont {K.~A.}\ \bibnamefont {M\"uller}},\
  }\bibfield  {title} {\bibinfo {title} {{Preparation and characterization of
  alkaline-earth substituted superconducting La$_2$CuO$_4$}},\ }\href
  {https://doi.org/https://doi.org/10.1016/0025-5408(87)90037-7} {\bibfield
  {journal} {\bibinfo  {journal} {Mat. Res. Bull.}\ }\textbf {\bibinfo {volume}
  {22}},\ \bibinfo {pages} {819} (\bibinfo {year} {1987})}\BibitemShut
  {NoStop}%
\bibitem [{\citenamefont {Keimer}\ \emph {et~al.}(2015)\citenamefont {Keimer},
  \citenamefont {Kivelson}, \citenamefont {Norman}, \citenamefont {Uchida},\
  and\ \citenamefont {Zaanen}}]{keim15}%
  \BibitemOpen
  \bibfield  {author} {\bibinfo {author} {\bibfnamefont {B.}~\bibnamefont
  {Keimer}}, \bibinfo {author} {\bibfnamefont {S.~A.}\ \bibnamefont
  {Kivelson}}, \bibinfo {author} {\bibfnamefont {M.~R.}\ \bibnamefont
  {Norman}}, \bibinfo {author} {\bibfnamefont {S.}~\bibnamefont {Uchida}},\
  and\ \bibinfo {author} {\bibfnamefont {J.}~\bibnamefont {Zaanen}},\
  }\bibfield  {title} {\bibinfo {title} {{From quantum matter to
  high-temperature superconductivity in copper oxides}},\ }\href
  {https://doi.org/10.1038/nature14165} {\bibfield  {journal} {\bibinfo
  {journal} {Nature}\ }\textbf {\bibinfo {volume} {518}},\ \bibinfo {pages}
  {179} (\bibinfo {year} {2015})}\BibitemShut {NoStop}%
\bibitem [{\citenamefont {Grissonnanche}\ \emph {et~al.}(2019)\citenamefont
  {Grissonnanche}, \citenamefont {Legros}, \citenamefont {Badoux},
  \citenamefont {Lefran\c{c}ois}, \citenamefont {Zatko}, \citenamefont
  {Lizaire}, \citenamefont {Lalibert\'e}, \citenamefont {Gourgout},
  \citenamefont {Zhou}, \citenamefont {Pyon}, \citenamefont {Takayama},
  \citenamefont {Takagi}, \citenamefont {Ono}, \citenamefont {Doiron-Leyraud},\
  and\ \citenamefont {Taillefer}}]{gris19}%
  \BibitemOpen
  \bibfield  {author} {\bibinfo {author} {\bibfnamefont {G.}~\bibnamefont
  {Grissonnanche}}, \bibinfo {author} {\bibfnamefont {A.}~\bibnamefont
  {Legros}}, \bibinfo {author} {\bibfnamefont {S.}~\bibnamefont {Badoux}},
  \bibinfo {author} {\bibfnamefont {E.}~\bibnamefont {Lefran\c{c}ois}},
  \bibinfo {author} {\bibfnamefont {V.}~\bibnamefont {Zatko}}, \bibinfo
  {author} {\bibfnamefont {M.}~\bibnamefont {Lizaire}}, \bibinfo {author}
  {\bibfnamefont {F.}~\bibnamefont {Lalibert\'e}}, \bibinfo {author}
  {\bibfnamefont {A.}~\bibnamefont {Gourgout}}, \bibinfo {author}
  {\bibfnamefont {J.~S.}\ \bibnamefont {Zhou}}, \bibinfo {author}
  {\bibfnamefont {S.}~\bibnamefont {Pyon}}, \bibinfo {author} {\bibfnamefont
  {T.}~\bibnamefont {Takayama}}, \bibinfo {author} {\bibfnamefont
  {H.}~\bibnamefont {Takagi}}, \bibinfo {author} {\bibfnamefont
  {S.}~\bibnamefont {Ono}}, \bibinfo {author} {\bibfnamefont {N.}~\bibnamefont
  {Doiron-Leyraud}},\ and\ \bibinfo {author} {\bibfnamefont {L.}~\bibnamefont
  {Taillefer}},\ }\bibfield  {title} {\bibinfo {title} {{Giant thermal Hall
  conductivity in the pseudogap phase of cuprate superconductors}},\ }\href
  {https://doi.org/10.1038/s41586-019-1375-0} {\bibfield  {journal} {\bibinfo
  {journal} {Nature}\ }\textbf {\bibinfo {volume} {571}},\ \bibinfo {pages}
  {376} (\bibinfo {year} {2019})}\BibitemShut {NoStop}%
\bibitem [{\citenamefont {Kastner}\ \emph {et~al.}(1988)\citenamefont
  {Kastner}, \citenamefont {Birgeneau}, \citenamefont {Thurston}, \citenamefont
  {Picone}, \citenamefont {Jenssen}, \citenamefont {Gabbe}, \citenamefont
  {Sato}, \citenamefont {Fukuda}, \citenamefont {Shamoto}, \citenamefont
  {Endoh}, \citenamefont {Yamada},\ and\ \citenamefont {Shirane}}]{kast88}%
  \BibitemOpen
  \bibfield  {author} {\bibinfo {author} {\bibfnamefont {M.~A.}\ \bibnamefont
  {Kastner}}, \bibinfo {author} {\bibfnamefont {R.~J.}\ \bibnamefont
  {Birgeneau}}, \bibinfo {author} {\bibfnamefont {T.~R.}\ \bibnamefont
  {Thurston}}, \bibinfo {author} {\bibfnamefont {P.~J.}\ \bibnamefont
  {Picone}}, \bibinfo {author} {\bibfnamefont {H.~P.}\ \bibnamefont {Jenssen}},
  \bibinfo {author} {\bibfnamefont {D.~R.}\ \bibnamefont {Gabbe}}, \bibinfo
  {author} {\bibfnamefont {M.}~\bibnamefont {Sato}}, \bibinfo {author}
  {\bibfnamefont {K.}~\bibnamefont {Fukuda}}, \bibinfo {author} {\bibfnamefont
  {S.}~\bibnamefont {Shamoto}}, \bibinfo {author} {\bibfnamefont
  {Y.}~\bibnamefont {Endoh}}, \bibinfo {author} {\bibfnamefont
  {K.}~\bibnamefont {Yamada}},\ and\ \bibinfo {author} {\bibfnamefont
  {G.}~\bibnamefont {Shirane}},\ }\bibfield  {title} {\bibinfo {title}
  {{Neutron-scattering study of the transition from antiferromagnetic to weak
  ferromagnetic order in ${\mathrm{La}}_{2}$Cu${\mathrm{O}}_{4}$}},\ }\href
  {https://doi.org/10.1103/PhysRevB.38.6636} {\bibfield  {journal} {\bibinfo
  {journal} {Phys. Rev. B}\ }\textbf {\bibinfo {volume} {38}},\ \bibinfo
  {pages} {6636} (\bibinfo {year} {1988})}\BibitemShut {NoStop}%
\bibitem [{\citenamefont {Reehuis}\ \emph {et~al.}(2006)\citenamefont
  {Reehuis}, \citenamefont {Ulrich}, \citenamefont {Proke\v{s}}, \citenamefont
  {Gozar}, \citenamefont {Blumberg}, \citenamefont {Komiya}, \citenamefont
  {Ando}, \citenamefont {Pattison},\ and\ \citenamefont {Keimer}}]{reeh06}%
  \BibitemOpen
  \bibfield  {author} {\bibinfo {author} {\bibfnamefont {M.}~\bibnamefont
  {Reehuis}}, \bibinfo {author} {\bibfnamefont {C.}~\bibnamefont {Ulrich}},
  \bibinfo {author} {\bibfnamefont {K.}~\bibnamefont {Proke\v{s}}}, \bibinfo
  {author} {\bibfnamefont {A.}~\bibnamefont {Gozar}}, \bibinfo {author}
  {\bibfnamefont {G.}~\bibnamefont {Blumberg}}, \bibinfo {author}
  {\bibfnamefont {S.}~\bibnamefont {Komiya}}, \bibinfo {author} {\bibfnamefont
  {Y.}~\bibnamefont {Ando}}, \bibinfo {author} {\bibfnamefont {P.}~\bibnamefont
  {Pattison}},\ and\ \bibinfo {author} {\bibfnamefont {B.}~\bibnamefont
  {Keimer}},\ }\bibfield  {title} {\bibinfo {title} {{Crystal structure and
  high-field magnetism of ${\mathrm{La}}_{2}\mathrm{Cu}{\mathrm{O}}_{4}$}},\
  }\href {https://doi.org/10.1103/PhysRevB.73.144513} {\bibfield  {journal}
  {\bibinfo  {journal} {Phys. Rev. B}\ }\textbf {\bibinfo {volume} {73}},\
  \bibinfo {pages} {144513} (\bibinfo {year} {2006})}\BibitemShut {NoStop}%
\bibitem [{\citenamefont {Grissonnanche}\ \emph {et~al.}(2020)\citenamefont
  {Grissonnanche}, \citenamefont {Th\'eriault}, \citenamefont {Gourgout},
  \citenamefont {Boulanger}, \citenamefont {Lefran\c{c}ois}, \citenamefont
  {Ataei}, \citenamefont {Lalibert\'e}, \citenamefont {Dion}, \citenamefont
  {Zhou}, \citenamefont {Pyon}, \citenamefont {Takayama}, \citenamefont
  {Takagi}, \citenamefont {Doiron-Leyraud},\ and\ \citenamefont
  {Taillefer}}]{gris20}%
  \BibitemOpen
  \bibfield  {author} {\bibinfo {author} {\bibfnamefont {G.}~\bibnamefont
  {Grissonnanche}}, \bibinfo {author} {\bibfnamefont {S.}~\bibnamefont
  {Th\'eriault}}, \bibinfo {author} {\bibfnamefont {A.}~\bibnamefont
  {Gourgout}}, \bibinfo {author} {\bibfnamefont {M.~E.}\ \bibnamefont
  {Boulanger}}, \bibinfo {author} {\bibfnamefont {E.}~\bibnamefont
  {Lefran\c{c}ois}}, \bibinfo {author} {\bibfnamefont {A.}~\bibnamefont
  {Ataei}}, \bibinfo {author} {\bibfnamefont {F.}~\bibnamefont {Lalibert\'e}},
  \bibinfo {author} {\bibfnamefont {M.}~\bibnamefont {Dion}}, \bibinfo {author}
  {\bibfnamefont {J.~S.}\ \bibnamefont {Zhou}}, \bibinfo {author}
  {\bibfnamefont {S.}~\bibnamefont {Pyon}}, \bibinfo {author} {\bibfnamefont
  {T.}~\bibnamefont {Takayama}}, \bibinfo {author} {\bibfnamefont
  {H.}~\bibnamefont {Takagi}}, \bibinfo {author} {\bibfnamefont
  {N.}~\bibnamefont {Doiron-Leyraud}},\ and\ \bibinfo {author} {\bibfnamefont
  {L.}~\bibnamefont {Taillefer}},\ }\bibfield  {title} {\bibinfo {title}
  {{Chiral phonons in the pseudogap phase of cuprates}},\ }\href
  {https://doi.org/10.1038/s41567-020-0965-y} {\bibfield  {journal} {\bibinfo
  {journal} {Nat. Phys.}\ }\textbf {\bibinfo {volume} {16}},\ \bibinfo {pages}
  {1108} (\bibinfo {year} {2020})}\BibitemShut {NoStop}%
\bibitem [{\citenamefont {Li}\ \emph {et~al.}(2020)\citenamefont {Li},
  \citenamefont {Fauqu\'e}, \citenamefont {Zhu},\ and\ \citenamefont
  {Behnia}}]{li20}%
  \BibitemOpen
  \bibfield  {author} {\bibinfo {author} {\bibfnamefont {X.}~\bibnamefont
  {Li}}, \bibinfo {author} {\bibfnamefont {B.}~\bibnamefont {Fauqu\'e}},
  \bibinfo {author} {\bibfnamefont {Z.}~\bibnamefont {Zhu}},\ and\ \bibinfo
  {author} {\bibfnamefont {K.}~\bibnamefont {Behnia}},\ }\bibfield  {title}
  {\bibinfo {title} {{Phonon Thermal Hall Effect in Strontium Titanate}},\
  }\href {https://doi.org/10.1103/PhysRevLett.124.105901} {\bibfield  {journal}
  {\bibinfo  {journal} {Phys. Rev. Lett.}\ }\textbf {\bibinfo {volume} {124}},\
  \bibinfo {pages} {105901} (\bibinfo {year} {2020})}\BibitemShut {NoStop}%
\bibitem [{\citenamefont {Samajdar}\ \emph {et~al.}(2019)\citenamefont
  {Samajdar}, \citenamefont {Scheurer}, \citenamefont {Chatterjee},
  \citenamefont {Guo}, \citenamefont {Xu},\ and\ \citenamefont
  {Sachdev}}]{sama19}%
  \BibitemOpen
  \bibfield  {author} {\bibinfo {author} {\bibfnamefont {R.}~\bibnamefont
  {Samajdar}}, \bibinfo {author} {\bibfnamefont {M.~S.}\ \bibnamefont
  {Scheurer}}, \bibinfo {author} {\bibfnamefont {S.}~\bibnamefont
  {Chatterjee}}, \bibinfo {author} {\bibfnamefont {H.}~\bibnamefont {Guo}},
  \bibinfo {author} {\bibfnamefont {C.}~\bibnamefont {Xu}},\ and\ \bibinfo
  {author} {\bibfnamefont {S.}~\bibnamefont {Sachdev}},\ }\bibfield  {title}
  {\bibinfo {title} {{Enhanced thermal Hall effect in the square-lattice N\'eel
  state}},\ }\href {https://doi.org/10.1038/s41567-019-0669-3} {\bibfield
  {journal} {\bibinfo  {journal} {Nat. Phys.}\ }\textbf {\bibinfo {volume}
  {15}},\ \bibinfo {pages} {1290} (\bibinfo {year} {2019})}\BibitemShut
  {NoStop}%
\bibitem [{\citenamefont {Han}\ \emph {et~al.}(2019)\citenamefont {Han},
  \citenamefont {Park},\ and\ \citenamefont {Lee}}]{han19}%
  \BibitemOpen
  \bibfield  {author} {\bibinfo {author} {\bibfnamefont {J.~H.}\ \bibnamefont
  {Han}}, \bibinfo {author} {\bibfnamefont {J.-H.}\ \bibnamefont {Park}},\ and\
  \bibinfo {author} {\bibfnamefont {P.~A.}\ \bibnamefont {Lee}},\ }\bibfield
  {title} {\bibinfo {title} {{Consideration of thermal Hall effect in undoped
  cuprates}},\ }\href {https://doi.org/10.1103/PhysRevB.99.205157} {\bibfield
  {journal} {\bibinfo  {journal} {Phys. Rev. B}\ }\textbf {\bibinfo {volume}
  {99}},\ \bibinfo {pages} {205157} (\bibinfo {year} {2019})}\BibitemShut
  {NoStop}%
\bibitem [{\citenamefont {Varma}(2020)}]{varm20}%
  \BibitemOpen
  \bibfield  {author} {\bibinfo {author} {\bibfnamefont {C.~M.}\ \bibnamefont
  {Varma}},\ }\bibfield  {title} {\bibinfo {title} {{Thermal Hall effect in the
  pseudogap phase of cuprates}},\ }\href
  {https://doi.org/10.1103/PhysRevB.102.075113} {\bibfield  {journal} {\bibinfo
   {journal} {Phys. Rev. B}\ }\textbf {\bibinfo {volume} {102}},\ \bibinfo
  {pages} {075113} (\bibinfo {year} {2020})}\BibitemShut {NoStop}%
\bibitem [{\citenamefont {Chen}\ \emph {et~al.}(2020)\citenamefont {Chen},
  \citenamefont {Kivelson},\ and\ \citenamefont {Sun}}]{chen20}%
  \BibitemOpen
  \bibfield  {author} {\bibinfo {author} {\bibfnamefont {J.-Y.}\ \bibnamefont
  {Chen}}, \bibinfo {author} {\bibfnamefont {S.~A.}\ \bibnamefont {Kivelson}},\
  and\ \bibinfo {author} {\bibfnamefont {X.-Q.}\ \bibnamefont {Sun}},\
  }\bibfield  {title} {\bibinfo {title} {{Enhanced Thermal Hall Effect in
  Nearly Ferroelectric Insulators}},\ }\href
  {https://doi.org/10.1103/PhysRevLett.124.167601} {\bibfield  {journal}
  {\bibinfo  {journal} {Phys. Rev. Lett.}\ }\textbf {\bibinfo {volume} {124}},\
  \bibinfo {pages} {167601} (\bibinfo {year} {2020})}\BibitemShut {NoStop}%
\bibitem [{\citenamefont {Guo}\ and\ \citenamefont {Sachdev}(2021)}]{guo21}%
  \BibitemOpen
  \bibfield  {author} {\bibinfo {author} {\bibfnamefont {H.}~\bibnamefont
  {Guo}}\ and\ \bibinfo {author} {\bibfnamefont {S.}~\bibnamefont {Sachdev}},\
  }\bibfield  {title} {\bibinfo {title} {{Extrinsic phonon thermal Hall
  transport from Hall viscosity}},\ }\href
  {https://doi.org/10.1103/PhysRevB.103.205115} {\bibfield  {journal} {\bibinfo
   {journal} {Phys. Rev. B}\ }\textbf {\bibinfo {volume} {103}},\ \bibinfo
  {pages} {205115} (\bibinfo {year} {2021})}\BibitemShut {NoStop}%
\bibitem [{\citenamefont {Boulanger}\ \emph {et~al.}(2020)\citenamefont
  {Boulanger}, \citenamefont {Grissonnanche}, \citenamefont {Badoux},
  \citenamefont {Allaire}, \citenamefont {Lefran{\c c}ois}, \citenamefont
  {Legros}, \citenamefont {Gourgout}, \citenamefont {Dion}, \citenamefont
  {Wang}, \citenamefont {Chen}, \citenamefont {Liang}, \citenamefont {Hardy},
  \citenamefont {Bonn},\ and\ \citenamefont {Taillefer}}]{boul20}%
  \BibitemOpen
  \bibfield  {author} {\bibinfo {author} {\bibfnamefont {M.-E.}\ \bibnamefont
  {Boulanger}}, \bibinfo {author} {\bibfnamefont {G.}~\bibnamefont
  {Grissonnanche}}, \bibinfo {author} {\bibfnamefont {S.}~\bibnamefont
  {Badoux}}, \bibinfo {author} {\bibfnamefont {A.}~\bibnamefont {Allaire}},
  \bibinfo {author} {\bibfnamefont {{\'E}.}~\bibnamefont {Lefran{\c c}ois}},
  \bibinfo {author} {\bibfnamefont {A.}~\bibnamefont {Legros}}, \bibinfo
  {author} {\bibfnamefont {A.}~\bibnamefont {Gourgout}}, \bibinfo {author}
  {\bibfnamefont {M.}~\bibnamefont {Dion}}, \bibinfo {author} {\bibfnamefont
  {C.~H.}\ \bibnamefont {Wang}}, \bibinfo {author} {\bibfnamefont {X.~H.}\
  \bibnamefont {Chen}}, \bibinfo {author} {\bibfnamefont {R.}~\bibnamefont
  {Liang}}, \bibinfo {author} {\bibfnamefont {W.~N.}\ \bibnamefont {Hardy}},
  \bibinfo {author} {\bibfnamefont {D.~A.}\ \bibnamefont {Bonn}},\ and\
  \bibinfo {author} {\bibfnamefont {L.}~\bibnamefont {Taillefer}},\ }\bibfield
  {title} {\bibinfo {title} {{Thermal Hall conductivity in the cuprate Mott
  insulators Nd$_2$CuO$_4$ and Sr$_2$CuO$_2$Cl$_2$}},\ }\href
  {https://doi.org/10.1038/s41467-020-18881-z} {\bibfield  {journal} {\bibinfo
  {journal} {Nat. Commun.}\ }\textbf {\bibinfo {volume} {11}},\ \bibinfo
  {pages} {5325} (\bibinfo {year} {2020})}\BibitemShut {NoStop}%
\bibitem [{\citenamefont {Sapkota}\ \emph {et~al.}(2021)\citenamefont
  {Sapkota}, \citenamefont {Sterling}, \citenamefont {Lozano}, \citenamefont
  {Li}, \citenamefont {Cao}, \citenamefont {Garlea}, \citenamefont {Reznik},
  \citenamefont {Li}, \citenamefont {Zaliznyak}, \citenamefont {Gu},\ and\
  \citenamefont {Tranquada}}]{sapk21}%
  \BibitemOpen
  \bibfield  {author} {\bibinfo {author} {\bibfnamefont {A.}~\bibnamefont
  {Sapkota}}, \bibinfo {author} {\bibfnamefont {T.~C.}\ \bibnamefont
  {Sterling}}, \bibinfo {author} {\bibfnamefont {P.~M.}\ \bibnamefont
  {Lozano}}, \bibinfo {author} {\bibfnamefont {Y.}~\bibnamefont {Li}}, \bibinfo
  {author} {\bibfnamefont {H.}~\bibnamefont {Cao}}, \bibinfo {author}
  {\bibfnamefont {V.~O.}\ \bibnamefont {Garlea}}, \bibinfo {author}
  {\bibfnamefont {D.}~\bibnamefont {Reznik}}, \bibinfo {author} {\bibfnamefont
  {Q.}~\bibnamefont {Li}}, \bibinfo {author} {\bibfnamefont {I.~A.}\
  \bibnamefont {Zaliznyak}}, \bibinfo {author} {\bibfnamefont {G.~D.}\
  \bibnamefont {Gu}},\ and\ \bibinfo {author} {\bibfnamefont {J.~M.}\
  \bibnamefont {Tranquada}},\ }\bibfield  {title} {\bibinfo {title}
  {{Reinvestigation of crystal symmetry and fluctuations in
  ${\mathrm{La}}_{2}{\mathrm{CuO}}_{4}$}},\ }\href
  {https://doi.org/10.1103/PhysRevB.104.014304} {\bibfield  {journal} {\bibinfo
   {journal} {Phys. Rev. B}\ }\textbf {\bibinfo {volume} {104}},\ \bibinfo
  {pages} {014304} (\bibinfo {year} {2021})}\BibitemShut {NoStop}%
\bibitem [{\citenamefont {Lehuede}\ and\ \citenamefont {Daire}(1973)}]{lehu73}%
  \BibitemOpen
  \bibfield  {author} {\bibinfo {author} {\bibfnamefont {P.}~\bibnamefont
  {Lehuede}}\ and\ \bibinfo {author} {\bibfnamefont {M.}~\bibnamefont
  {Daire}},\ }\bibfield  {title} {\bibinfo {title} {{Sur la structure et les
  propri\'et\'es du compos\'e La$_2$CuO$_4$}},\ }\href@noop {} {\bibfield
  {journal} {\bibinfo  {journal} {C. R. Acad. Sci. Paris}\ }\textbf {\bibinfo
  {volume} {276}},\ \bibinfo {pages} {{C}} (\bibinfo {year}
  {1973})}\BibitemShut {NoStop}%
\bibitem [{\citenamefont {Longo}\ and\ \citenamefont {Raccah}(1973)}]{long73}%
  \BibitemOpen
  \bibfield  {author} {\bibinfo {author} {\bibfnamefont {J.~M.}\ \bibnamefont
  {Longo}}\ and\ \bibinfo {author} {\bibfnamefont {P.~M.}\ \bibnamefont
  {Raccah}},\ }\bibfield  {title} {\bibinfo {title} {{The structure of
  La$_2$CuO$_4$ and LaSrVO$_4$}},\ }\href
  {https://doi.org/https://doi.org/10.1016/S0022-4596(73)80010-6} {\bibfield
  {journal} {\bibinfo  {journal} {J. Solid State Chem.}\ }\textbf {\bibinfo
  {volume} {6}},\ \bibinfo {pages} {526} (\bibinfo {year} {1973})}\BibitemShut
  {NoStop}%
\bibitem [{\citenamefont {H\"ucker}\ \emph {et~al.}(2004)\citenamefont
  {H\"ucker}, \citenamefont {Kataev}, \citenamefont {Pommer}, \citenamefont
  {Ammerahl}, \citenamefont {Revcolevschi}, \citenamefont {Tranquada},\ and\
  \citenamefont {B\"uchner}}]{huck04}%
  \BibitemOpen
  \bibfield  {author} {\bibinfo {author} {\bibfnamefont {M.}~\bibnamefont
  {H\"ucker}}, \bibinfo {author} {\bibfnamefont {V.}~\bibnamefont {Kataev}},
  \bibinfo {author} {\bibfnamefont {J.}~\bibnamefont {Pommer}}, \bibinfo
  {author} {\bibfnamefont {U.}~\bibnamefont {Ammerahl}}, \bibinfo {author}
  {\bibfnamefont {A.}~\bibnamefont {Revcolevschi}}, \bibinfo {author}
  {\bibfnamefont {J.~M.}\ \bibnamefont {Tranquada}},\ and\ \bibinfo {author}
  {\bibfnamefont {B.}~\bibnamefont {B\"uchner}},\ }\bibfield  {title} {\bibinfo
  {title} {{Dzyaloshinsky-Moriya spin canting in the low-temperature tetragonal
  phase of
  ${\mathrm{La}}_{2\ensuremath{-}x\ensuremath{-}y}{\mathrm{Eu}}_{y}{\mathrm{Sr}}_{x}\mathrm{Cu}{\mathrm{O}}_{4}$}},\
  }\href {https://doi.org/10.1103/PhysRevB.70.214515} {\bibfield  {journal}
  {\bibinfo  {journal} {Phys. Rev. B}\ }\textbf {\bibinfo {volume} {70}},\
  \bibinfo {pages} {214515} (\bibinfo {year} {2004})}\BibitemShut {NoStop}%
\bibitem [{\citenamefont {Tokura}\ \emph {et~al.}(1990)\citenamefont {Tokura},
  \citenamefont {Koshihara}, \citenamefont {Arima}, \citenamefont {Takagi},
  \citenamefont {Ishibashi}, \citenamefont {Ido},\ and\ \citenamefont
  {Uchida}}]{toku90}%
  \BibitemOpen
  \bibfield  {author} {\bibinfo {author} {\bibfnamefont {Y.}~\bibnamefont
  {Tokura}}, \bibinfo {author} {\bibfnamefont {S.}~\bibnamefont {Koshihara}},
  \bibinfo {author} {\bibfnamefont {T.}~\bibnamefont {Arima}}, \bibinfo
  {author} {\bibfnamefont {H.}~\bibnamefont {Takagi}}, \bibinfo {author}
  {\bibfnamefont {S.}~\bibnamefont {Ishibashi}}, \bibinfo {author}
  {\bibfnamefont {T.}~\bibnamefont {Ido}},\ and\ \bibinfo {author}
  {\bibfnamefont {S.}~\bibnamefont {Uchida}},\ }\bibfield  {title} {\bibinfo
  {title} {{Cu-O network dependence of optical charge-transfer gaps and
  spin-pair excitations in single-${\mathrm{CuO}}_{2}$-layer compounds}},\
  }\href {https://doi.org/10.1103/PhysRevB.41.11657} {\bibfield  {journal}
  {\bibinfo  {journal} {Phys. Rev. B}\ }\textbf {\bibinfo {volume} {41}},\
  \bibinfo {pages} {11657} (\bibinfo {year} {1990})}\BibitemShut {NoStop}%
\bibitem [{\citenamefont {Zaanen}\ \emph {et~al.}(1985)\citenamefont {Zaanen},
  \citenamefont {Sawatzky},\ and\ \citenamefont {Allen}}]{zaan85}%
  \BibitemOpen
  \bibfield  {author} {\bibinfo {author} {\bibfnamefont {J.}~\bibnamefont
  {Zaanen}}, \bibinfo {author} {\bibfnamefont {G.~A.}\ \bibnamefont
  {Sawatzky}},\ and\ \bibinfo {author} {\bibfnamefont {J.~W.}\ \bibnamefont
  {Allen}},\ }\bibfield  {title} {\bibinfo {title} {{Band gaps and electronic
  structure of transition-metal compounds}},\ }\href
  {https://doi.org/10.1103/PhysRevLett.55.418} {\bibfield  {journal} {\bibinfo
  {journal} {Phys. Rev. Lett.}\ }\textbf {\bibinfo {volume} {55}},\ \bibinfo
  {pages} {418} (\bibinfo {year} {1985})}\BibitemShut {NoStop}%
\bibitem [{\citenamefont {Shirane}\ \emph {et~al.}(1987)\citenamefont
  {Shirane}, \citenamefont {Endoh}, \citenamefont {Birgeneau}, \citenamefont
  {Kastner}, \citenamefont {Hidaka}, \citenamefont {Oda}, \citenamefont
  {Suzuki},\ and\ \citenamefont {Murakami}}]{shir87}%
  \BibitemOpen
  \bibfield  {author} {\bibinfo {author} {\bibfnamefont {G.}~\bibnamefont
  {Shirane}}, \bibinfo {author} {\bibfnamefont {Y.}~\bibnamefont {Endoh}},
  \bibinfo {author} {\bibfnamefont {R.~J.}\ \bibnamefont {Birgeneau}}, \bibinfo
  {author} {\bibfnamefont {M.~A.}\ \bibnamefont {Kastner}}, \bibinfo {author}
  {\bibfnamefont {Y.}~\bibnamefont {Hidaka}}, \bibinfo {author} {\bibfnamefont
  {M.}~\bibnamefont {Oda}}, \bibinfo {author} {\bibfnamefont {M.}~\bibnamefont
  {Suzuki}},\ and\ \bibinfo {author} {\bibfnamefont {T.}~\bibnamefont
  {Murakami}},\ }\bibfield  {title} {\bibinfo {title} {{Two-dimensional
  antiferromagnetic quantum spin-fluid state in
  ${\mathrm{La}}_{2}$${\mathrm{CuO}}_{4}$}},\ }\href
  {https://doi.org/10.1103/PhysRevLett.59.1613} {\bibfield  {journal} {\bibinfo
   {journal} {Phys. Rev. Lett.}\ }\textbf {\bibinfo {volume} {59}},\ \bibinfo
  {pages} {1613} (\bibinfo {year} {1987})}\BibitemShut {NoStop}%
\bibitem [{\citenamefont {Imai}\ \emph {et~al.}(1993)\citenamefont {Imai},
  \citenamefont {Slichter}, \citenamefont {Yoshimura},\ and\ \citenamefont
  {Kosuge}}]{imai93}%
  \BibitemOpen
  \bibfield  {author} {\bibinfo {author} {\bibfnamefont {T.}~\bibnamefont
  {Imai}}, \bibinfo {author} {\bibfnamefont {C.~P.}\ \bibnamefont {Slichter}},
  \bibinfo {author} {\bibfnamefont {K.}~\bibnamefont {Yoshimura}},\ and\
  \bibinfo {author} {\bibfnamefont {K.}~\bibnamefont {Kosuge}},\ }\bibfield
  {title} {\bibinfo {title} {{Low frequency spin dynamics in undoped and
  Sr-doped ${\mathrm{La}}_{2}$${\mathrm{CuO}}_{4}$}},\ }\href
  {https://doi.org/10.1103/PhysRevLett.70.1002} {\bibfield  {journal} {\bibinfo
   {journal} {Phys. Rev. Lett.}\ }\textbf {\bibinfo {volume} {70}},\ \bibinfo
  {pages} {1002} (\bibinfo {year} {1993})}\BibitemShut {NoStop}%
\bibitem [{\citenamefont {Birgeneau}\ \emph {et~al.}(1999)\citenamefont
  {Birgeneau}, \citenamefont {Greven}, \citenamefont {Kastner}, \citenamefont
  {Lee}, \citenamefont {Wells}, \citenamefont {Endoh}, \citenamefont {Yamada},\
  and\ \citenamefont {Shirane}}]{birg99}%
  \BibitemOpen
  \bibfield  {author} {\bibinfo {author} {\bibfnamefont {R.~J.}\ \bibnamefont
  {Birgeneau}}, \bibinfo {author} {\bibfnamefont {M.}~\bibnamefont {Greven}},
  \bibinfo {author} {\bibfnamefont {M.~A.}\ \bibnamefont {Kastner}}, \bibinfo
  {author} {\bibfnamefont {Y.~S.}\ \bibnamefont {Lee}}, \bibinfo {author}
  {\bibfnamefont {B.~O.}\ \bibnamefont {Wells}}, \bibinfo {author}
  {\bibfnamefont {Y.}~\bibnamefont {Endoh}}, \bibinfo {author} {\bibfnamefont
  {K.}~\bibnamefont {Yamada}},\ and\ \bibinfo {author} {\bibfnamefont
  {G.}~\bibnamefont {Shirane}},\ }\bibfield  {title} {\bibinfo {title}
  {{Instantaneous spin correlations in
  ${\mathrm{La}}_{2}{\mathrm{CuO}}_{4}$}},\ }\href
  {https://doi.org/10.1103/PhysRevB.59.13788} {\bibfield  {journal} {\bibinfo
  {journal} {Phys. Rev. B}\ }\textbf {\bibinfo {volume} {59}},\ \bibinfo
  {pages} {13788} (\bibinfo {year} {1999})}\BibitemShut {NoStop}%
\bibitem [{\citenamefont {Anderson}(1987)}]{ande87}%
  \BibitemOpen
  \bibfield  {author} {\bibinfo {author} {\bibfnamefont {P.~W.}\ \bibnamefont
  {Anderson}},\ }\bibfield  {title} {\bibinfo {title} {{The Resonating Valence
  Bond State in La$_2$CuO$_4$ and Superconductivity}},\ }\href
  {https://doi.org/10.1126/science.235.4793.1196} {\bibfield  {journal}
  {\bibinfo  {journal} {Science}\ }\textbf {\bibinfo {volume} {235}},\ \bibinfo
  {pages} {1196} (\bibinfo {year} {1987})}\BibitemShut {NoStop}%
\bibitem [{\citenamefont {Coldea}\ \emph {et~al.}(2001)\citenamefont {Coldea},
  \citenamefont {Hayden}, \citenamefont {Aeppli}, \citenamefont {Perring},
  \citenamefont {Frost}, \citenamefont {Mason}, \citenamefont {Cheong},\ and\
  \citenamefont {Fisk}}]{cold01}%
  \BibitemOpen
  \bibfield  {author} {\bibinfo {author} {\bibfnamefont {R.}~\bibnamefont
  {Coldea}}, \bibinfo {author} {\bibfnamefont {S.~M.}\ \bibnamefont {Hayden}},
  \bibinfo {author} {\bibfnamefont {G.}~\bibnamefont {Aeppli}}, \bibinfo
  {author} {\bibfnamefont {T.~G.}\ \bibnamefont {Perring}}, \bibinfo {author}
  {\bibfnamefont {C.~D.}\ \bibnamefont {Frost}}, \bibinfo {author}
  {\bibfnamefont {T.~E.}\ \bibnamefont {Mason}}, \bibinfo {author}
  {\bibfnamefont {S.-W.}\ \bibnamefont {Cheong}},\ and\ \bibinfo {author}
  {\bibfnamefont {Z.}~\bibnamefont {Fisk}},\ }\bibfield  {title} {\bibinfo
  {title} {{Spin Waves and Electronic Interactions in
  ${\mathrm{La}}_{2}{\mathrm{CuO}}_{4}$}},\ }\href
  {https://doi.org/10.1103/PhysRevLett.86.5377} {\bibfield  {journal} {\bibinfo
   {journal} {Phys. Rev. Lett.}\ }\textbf {\bibinfo {volume} {86}},\ \bibinfo
  {pages} {5377} (\bibinfo {year} {2001})}\BibitemShut {NoStop}%
\bibitem [{\citenamefont {Vaknin}\ \emph {et~al.}(1987)\citenamefont {Vaknin},
  \citenamefont {Sinha}, \citenamefont {Moncton}, \citenamefont {Johnston},
  \citenamefont {Newsam}, \citenamefont {Safinya},\ and\ \citenamefont
  {King}}]{vakn87}%
  \BibitemOpen
  \bibfield  {author} {\bibinfo {author} {\bibfnamefont {D.}~\bibnamefont
  {Vaknin}}, \bibinfo {author} {\bibfnamefont {S.~K.}\ \bibnamefont {Sinha}},
  \bibinfo {author} {\bibfnamefont {D.~E.}\ \bibnamefont {Moncton}}, \bibinfo
  {author} {\bibfnamefont {D.~C.}\ \bibnamefont {Johnston}}, \bibinfo {author}
  {\bibfnamefont {J.~M.}\ \bibnamefont {Newsam}}, \bibinfo {author}
  {\bibfnamefont {C.~R.}\ \bibnamefont {Safinya}},\ and\ \bibinfo {author}
  {\bibfnamefont {H.~E.}\ \bibnamefont {King}},\ }\bibfield  {title} {\bibinfo
  {title} {{Antiferromagnetism in
  ${\mathrm{La}}_{2}$${\mathrm{CuO}}_{4\mathrm{\ensuremath{-}}\mathrm{y}}$}},\
  }\href {https://doi.org/10.1103/PhysRevLett.58.2802} {\bibfield  {journal}
  {\bibinfo  {journal} {Phys. Rev. Lett.}\ }\textbf {\bibinfo {volume} {58}},\
  \bibinfo {pages} {2802} (\bibinfo {year} {1987})}\BibitemShut {NoStop}%
\bibitem [{\citenamefont {Budnick}\ \emph {et~al.}(1987)\citenamefont
  {Budnick}, \citenamefont {Golnik}, \citenamefont {Niedermayer}, \citenamefont
  {Recknagel}, \citenamefont {Rossmanith}, \citenamefont {Weidinger},
  \citenamefont {Chamberland}, \citenamefont {Filipkowski},\ and\ \citenamefont
  {Yang}}]{budn87}%
  \BibitemOpen
  \bibfield  {author} {\bibinfo {author} {\bibfnamefont {J.}~\bibnamefont
  {Budnick}}, \bibinfo {author} {\bibfnamefont {A.}~\bibnamefont {Golnik}},
  \bibinfo {author} {\bibfnamefont {C.}~\bibnamefont {Niedermayer}}, \bibinfo
  {author} {\bibfnamefont {E.}~\bibnamefont {Recknagel}}, \bibinfo {author}
  {\bibfnamefont {M.}~\bibnamefont {Rossmanith}}, \bibinfo {author}
  {\bibfnamefont {A.}~\bibnamefont {Weidinger}}, \bibinfo {author}
  {\bibfnamefont {B.}~\bibnamefont {Chamberland}}, \bibinfo {author}
  {\bibfnamefont {M.}~\bibnamefont {Filipkowski}},\ and\ \bibinfo {author}
  {\bibfnamefont {D.}~\bibnamefont {Yang}},\ }\bibfield  {title} {\bibinfo
  {title} {{Observation of magnetic ordering in La2CuO4 by muon spin rotation
  spectroscopy}},\ }\href@noop {} {\bibfield  {journal} {\bibinfo  {journal}
  {Phys. Lett. A}\ }\textbf {\bibinfo {volume} {124}},\ \bibinfo {pages} {103}
  (\bibinfo {year} {1987})}\BibitemShut {NoStop}%
\bibitem [{\citenamefont {Keimer}\ \emph {et~al.}(1992)\citenamefont {Keimer},
  \citenamefont {Aharony}, \citenamefont {Auerbach}, \citenamefont {Birgeneau},
  \citenamefont {Cassanho}, \citenamefont {Endoh}, \citenamefont {Erwin},
  \citenamefont {Kastner},\ and\ \citenamefont {Shirane}}]{keim92}%
  \BibitemOpen
  \bibfield  {author} {\bibinfo {author} {\bibfnamefont {B.}~\bibnamefont
  {Keimer}}, \bibinfo {author} {\bibfnamefont {A.}~\bibnamefont {Aharony}},
  \bibinfo {author} {\bibfnamefont {A.}~\bibnamefont {Auerbach}}, \bibinfo
  {author} {\bibfnamefont {R.~J.}\ \bibnamefont {Birgeneau}}, \bibinfo {author}
  {\bibfnamefont {A.}~\bibnamefont {Cassanho}}, \bibinfo {author}
  {\bibfnamefont {Y.}~\bibnamefont {Endoh}}, \bibinfo {author} {\bibfnamefont
  {R.~W.}\ \bibnamefont {Erwin}}, \bibinfo {author} {\bibfnamefont {M.~A.}\
  \bibnamefont {Kastner}},\ and\ \bibinfo {author} {\bibfnamefont
  {G.}~\bibnamefont {Shirane}},\ }\bibfield  {title} {\bibinfo {title} {{N\'eel
  transition and sublattice magnetization of pure and doped
  ${\mathrm{La}}_{2}$${\mathrm{CuO}}_{4}$}},\ }\href
  {https://doi.org/10.1103/PhysRevB.45.7430} {\bibfield  {journal} {\bibinfo
  {journal} {Phys. Rev. B}\ }\textbf {\bibinfo {volume} {45}},\ \bibinfo
  {pages} {7430} (\bibinfo {year} {1992})}\BibitemShut {NoStop}%
\bibitem [{\citenamefont {Wells}\ \emph {et~al.}(1997)\citenamefont {Wells},
  \citenamefont {Lee}, \citenamefont {Kastner}, \citenamefont {Christianson},
  \citenamefont {Birgeneau}, \citenamefont {Yamada}, \citenamefont {Endoh},\
  and\ \citenamefont {Shirane}}]{well97}%
  \BibitemOpen
  \bibfield  {author} {\bibinfo {author} {\bibfnamefont {B.~O.}\ \bibnamefont
  {Wells}}, \bibinfo {author} {\bibfnamefont {Y.~S.}\ \bibnamefont {Lee}},
  \bibinfo {author} {\bibfnamefont {M.~A.}\ \bibnamefont {Kastner}}, \bibinfo
  {author} {\bibfnamefont {R.~J.}\ \bibnamefont {Christianson}}, \bibinfo
  {author} {\bibfnamefont {R.~J.}\ \bibnamefont {Birgeneau}}, \bibinfo {author}
  {\bibfnamefont {K.}~\bibnamefont {Yamada}}, \bibinfo {author} {\bibfnamefont
  {Y.}~\bibnamefont {Endoh}},\ and\ \bibinfo {author} {\bibfnamefont
  {G.}~\bibnamefont {Shirane}},\ }\bibfield  {title} {\bibinfo {title}
  {{Incommensurate Spin Fluctuations in High-Transition Temperature
  Superconductors}},\ }\href {https://doi.org/10.1126/science.277.5329.1067}
  {\bibfield  {journal} {\bibinfo  {journal} {Science}\ }\textbf {\bibinfo
  {volume} {277}},\ \bibinfo {pages} {1067} (\bibinfo {year}
  {1997})}\BibitemShut {NoStop}%
\bibitem [{\citenamefont {Thio}\ \emph {et~al.}(1988)\citenamefont {Thio},
  \citenamefont {Thurston}, \citenamefont {Preyer}, \citenamefont {Picone},
  \citenamefont {Kastner}, \citenamefont {Jenssen}, \citenamefont {Gabbe},
  \citenamefont {Chen}, \citenamefont {Birgeneau},\ and\ \citenamefont
  {Aharony}}]{thio88}%
  \BibitemOpen
  \bibfield  {author} {\bibinfo {author} {\bibfnamefont {T.}~\bibnamefont
  {Thio}}, \bibinfo {author} {\bibfnamefont {T.~R.}\ \bibnamefont {Thurston}},
  \bibinfo {author} {\bibfnamefont {N.~W.}\ \bibnamefont {Preyer}}, \bibinfo
  {author} {\bibfnamefont {P.~J.}\ \bibnamefont {Picone}}, \bibinfo {author}
  {\bibfnamefont {M.~A.}\ \bibnamefont {Kastner}}, \bibinfo {author}
  {\bibfnamefont {H.~P.}\ \bibnamefont {Jenssen}}, \bibinfo {author}
  {\bibfnamefont {D.~R.}\ \bibnamefont {Gabbe}}, \bibinfo {author}
  {\bibfnamefont {C.~Y.}\ \bibnamefont {Chen}}, \bibinfo {author}
  {\bibfnamefont {R.~J.}\ \bibnamefont {Birgeneau}},\ and\ \bibinfo {author}
  {\bibfnamefont {A.}~\bibnamefont {Aharony}},\ }\bibfield  {title} {\bibinfo
  {title} {{Antisymmetric exchange and its influence on the magnetic structure
  and conductivity of ${\mathrm{La}}_{2}$Cu${\mathrm{O}}_{4}$}},\ }\href
  {https://doi.org/10.1103/PhysRevB.38.905} {\bibfield  {journal} {\bibinfo
  {journal} {Phys. Rev. B}\ }\textbf {\bibinfo {volume} {38}},\ \bibinfo
  {pages} {905} (\bibinfo {year} {1988})}\BibitemShut {NoStop}%
\bibitem [{\citenamefont {Lavrov}\ \emph {et~al.}(2001)\citenamefont {Lavrov},
  \citenamefont {Ando}, \citenamefont {Komiya},\ and\ \citenamefont
  {Tsukada}}]{lavr01}%
  \BibitemOpen
  \bibfield  {author} {\bibinfo {author} {\bibfnamefont {A.~N.}\ \bibnamefont
  {Lavrov}}, \bibinfo {author} {\bibfnamefont {Y.}~\bibnamefont {Ando}},
  \bibinfo {author} {\bibfnamefont {S.}~\bibnamefont {Komiya}},\ and\ \bibinfo
  {author} {\bibfnamefont {I.}~\bibnamefont {Tsukada}},\ }\bibfield  {title}
  {\bibinfo {title} {{Unusual Magnetic Susceptibility Anisotropy in Untwinned
  ${\mathrm{La}}_{2\ensuremath{-}\mathit{x}}{\mathrm{Sr}}_{\mathit{x}}{\mathrm{CuO}}_{4}$
  Single Crystals in the Lightly Doped Region}},\ }\href
  {https://doi.org/10.1103/PhysRevLett.87.017007} {\bibfield  {journal}
  {\bibinfo  {journal} {Phys. Rev. Lett.}\ }\textbf {\bibinfo {volume} {87}},\
  \bibinfo {pages} {017007} (\bibinfo {year} {2001})}\BibitemShut {NoStop}%
\bibitem [{\citenamefont {Dzyaloshinsky}(1958)}]{dzya58}%
  \BibitemOpen
  \bibfield  {author} {\bibinfo {author} {\bibfnamefont {I.}~\bibnamefont
  {Dzyaloshinsky}},\ }\bibfield  {title} {\bibinfo {title} {{A thermodynamic
  theory of ``weak'' ferromagnetism of antiferromagnetics}},\ }\href
  {https://doi.org/https://doi.org/10.1016/0022-3697(58)90076-3} {\bibfield
  {journal} {\bibinfo  {journal} {J. Phys. Chem. Solids}\ }\textbf {\bibinfo
  {volume} {4}},\ \bibinfo {pages} {241} (\bibinfo {year} {1958})}\BibitemShut
  {NoStop}%
\bibitem [{\citenamefont {Moriya}(1960)}]{mori60}%
  \BibitemOpen
  \bibfield  {author} {\bibinfo {author} {\bibfnamefont {T.}~\bibnamefont
  {Moriya}},\ }\bibfield  {title} {\bibinfo {title} {{Anisotropic Superexchange
  Interaction and Weak Ferromagnetism}},\ }\href
  {https://doi.org/10.1103/PhysRev.120.91} {\bibfield  {journal} {\bibinfo
  {journal} {Phys. Rev.}\ }\textbf {\bibinfo {volume} {120}},\ \bibinfo {pages}
  {91} (\bibinfo {year} {1960})}\BibitemShut {NoStop}%
\bibitem [{\citenamefont {Coffey}\ \emph {et~al.}(1991)\citenamefont {Coffey},
  \citenamefont {Rice},\ and\ \citenamefont {Zhang}}]{coff91}%
  \BibitemOpen
  \bibfield  {author} {\bibinfo {author} {\bibfnamefont {D.}~\bibnamefont
  {Coffey}}, \bibinfo {author} {\bibfnamefont {T.~M.}\ \bibnamefont {Rice}},\
  and\ \bibinfo {author} {\bibfnamefont {F.~C.}\ \bibnamefont {Zhang}},\
  }\bibfield  {title} {\bibinfo {title} {{Dzyaloshinskii-Moriya interaction in
  the cuprates}},\ }\href {https://doi.org/10.1103/PhysRevB.44.10112}
  {\bibfield  {journal} {\bibinfo  {journal} {Phys. Rev. B}\ }\textbf {\bibinfo
  {volume} {44}},\ \bibinfo {pages} {10112} (\bibinfo {year}
  {1991})}\BibitemShut {NoStop}%
\bibitem [{\citenamefont {Shekhtman}\ \emph {et~al.}(1992)\citenamefont
  {Shekhtman}, \citenamefont {Entin-Wohlman},\ and\ \citenamefont
  {Aharony}}]{shek92}%
  \BibitemOpen
  \bibfield  {author} {\bibinfo {author} {\bibfnamefont {L.}~\bibnamefont
  {Shekhtman}}, \bibinfo {author} {\bibfnamefont {O.}~\bibnamefont
  {Entin-Wohlman}},\ and\ \bibinfo {author} {\bibfnamefont {A.}~\bibnamefont
  {Aharony}},\ }\bibfield  {title} {\bibinfo {title} {{Moriya's anisotropic
  superexchange interaction, frustration, and Dzyaloshinsky's weak
  ferromagnetism}},\ }\href {https://doi.org/10.1103/PhysRevLett.69.836}
  {\bibfield  {journal} {\bibinfo  {journal} {Phys. Rev. Lett.}\ }\textbf
  {\bibinfo {volume} {69}},\ \bibinfo {pages} {836} (\bibinfo {year}
  {1992})}\BibitemShut {NoStop}%
\bibitem [{\citenamefont {Bonesteel}(1993)}]{bone93}%
  \BibitemOpen
  \bibfield  {author} {\bibinfo {author} {\bibfnamefont {N.~E.}\ \bibnamefont
  {Bonesteel}},\ }\bibfield  {title} {\bibinfo {title} {{Theory of anisotropic
  superexchange in insulating cuprates}},\ }\href
  {https://doi.org/10.1103/PhysRevB.47.11302} {\bibfield  {journal} {\bibinfo
  {journal} {Phys. Rev. B}\ }\textbf {\bibinfo {volume} {47}},\ \bibinfo
  {pages} {11302} (\bibinfo {year} {1993})}\BibitemShut {NoStop}%
\bibitem [{\citenamefont {Entin-Wohlman}\ \emph {et~al.}(1994)\citenamefont
  {Entin-Wohlman}, \citenamefont {Aharony},\ and\ \citenamefont
  {Shekhtman}}]{enti94}%
  \BibitemOpen
  \bibfield  {author} {\bibinfo {author} {\bibfnamefont {O.}~\bibnamefont
  {Entin-Wohlman}}, \bibinfo {author} {\bibfnamefont {A.}~\bibnamefont
  {Aharony}},\ and\ \bibinfo {author} {\bibfnamefont {L.}~\bibnamefont
  {Shekhtman}},\ }\bibfield  {title} {\bibinfo {title} {{Superexchange
  anisotropy in the cuprates}},\ }\href
  {https://doi.org/10.1103/PhysRevB.50.3068} {\bibfield  {journal} {\bibinfo
  {journal} {Phys. Rev. B}\ }\textbf {\bibinfo {volume} {50}},\ \bibinfo
  {pages} {3068} (\bibinfo {year} {1994})}\BibitemShut {NoStop}%
\bibitem [{\citenamefont {Koshibae}\ \emph {et~al.}(1994)\citenamefont
  {Koshibae}, \citenamefont {Ohta},\ and\ \citenamefont {Maekawa}}]{kosh94}%
  \BibitemOpen
  \bibfield  {author} {\bibinfo {author} {\bibfnamefont {W.}~\bibnamefont
  {Koshibae}}, \bibinfo {author} {\bibfnamefont {Y.}~\bibnamefont {Ohta}},\
  and\ \bibinfo {author} {\bibfnamefont {S.}~\bibnamefont {Maekawa}},\
  }\bibfield  {title} {\bibinfo {title} {{Theory of Dzyaloshinski-Moriya
  antiferromagnetism in distorted ${\mathrm{CuO}}_{2}$ and ${\mathrm{NiO}}_{2}$
  planes}},\ }\href {https://doi.org/10.1103/PhysRevB.50.3767} {\bibfield
  {journal} {\bibinfo  {journal} {Phys. Rev. B}\ }\textbf {\bibinfo {volume}
  {50}},\ \bibinfo {pages} {3767} (\bibinfo {year} {1994})}\BibitemShut
  {NoStop}%
\bibitem [{\citenamefont {Vierti\"o}\ and\ \citenamefont
  {Bonesteel}(1994)}]{vier94b}%
  \BibitemOpen
  \bibfield  {author} {\bibinfo {author} {\bibfnamefont {H.~E.}\ \bibnamefont
  {Vierti\"o}}\ and\ \bibinfo {author} {\bibfnamefont {N.~E.}\ \bibnamefont
  {Bonesteel}},\ }\bibfield  {title} {\bibinfo {title} {{Interplanar coupling
  and the weak ferromagnetic transition in
  ${\mathrm{La}}_{2\mathrm{\ensuremath{-}}\mathit{x}}$${\mathrm{Nd}}_{\mathit{x}}$${\mathrm{CuO}}_{4}$}},\
  }\href {https://doi.org/10.1103/PhysRevB.49.6088} {\bibfield  {journal}
  {\bibinfo  {journal} {Phys. Rev. B}\ }\textbf {\bibinfo {volume} {49}},\
  \bibinfo {pages} {6088} (\bibinfo {year} {1994})}\BibitemShut {NoStop}%
\bibitem [{\citenamefont {Yildirim}\ \emph
  {et~al.}(1994{\natexlab{a}})\citenamefont {Yildirim}, \citenamefont {Harris},
  \citenamefont {Entin-Wohlman},\ and\ \citenamefont {Aharony}}]{yild94b}%
  \BibitemOpen
  \bibfield  {author} {\bibinfo {author} {\bibfnamefont {T.}~\bibnamefont
  {Yildirim}}, \bibinfo {author} {\bibfnamefont {A.~B.}\ \bibnamefont
  {Harris}}, \bibinfo {author} {\bibfnamefont {O.}~\bibnamefont
  {Entin-Wohlman}},\ and\ \bibinfo {author} {\bibfnamefont {A.}~\bibnamefont
  {Aharony}},\ }\bibfield  {title} {\bibinfo {title} {{Symmetry, Spin-Orbit
  Interactions, and Spin Anisotropies}},\ }\href
  {https://doi.org/10.1103/PhysRevLett.73.2919} {\bibfield  {journal} {\bibinfo
   {journal} {Phys. Rev. Lett.}\ }\textbf {\bibinfo {volume} {73}},\ \bibinfo
  {pages} {2919} (\bibinfo {year} {1994}{\natexlab{a}})}\BibitemShut {NoStop}%
\bibitem [{\citenamefont {Stein}\ \emph {et~al.}(1996)\citenamefont {Stein},
  \citenamefont {Entin-Wohlman},\ and\ \citenamefont {Aharony}}]{stei96}%
  \BibitemOpen
  \bibfield  {author} {\bibinfo {author} {\bibfnamefont {J.}~\bibnamefont
  {Stein}}, \bibinfo {author} {\bibfnamefont {O.}~\bibnamefont
  {Entin-Wohlman}},\ and\ \bibinfo {author} {\bibfnamefont {A.}~\bibnamefont
  {Aharony}},\ }\bibfield  {title} {\bibinfo {title} {{Weak ferromagnetism in
  the low-temperature tetragonal phase of the cuprates}},\ }\href
  {https://doi.org/10.1103/PhysRevB.53.775} {\bibfield  {journal} {\bibinfo
  {journal} {Phys. Rev. B}\ }\textbf {\bibinfo {volume} {53}},\ \bibinfo
  {pages} {775} (\bibinfo {year} {1996})}\BibitemShut {NoStop}%
\bibitem [{\citenamefont {Keimer}\ \emph {et~al.}(1993)\citenamefont {Keimer},
  \citenamefont {Birgeneau}, \citenamefont {Cassanho}, \citenamefont {Endoh},
  \citenamefont {Greven}, \citenamefont {Kastner},\ and\ \citenamefont
  {Shirane}}]{keim93}%
  \BibitemOpen
  \bibfield  {author} {\bibinfo {author} {\bibfnamefont {B.}~\bibnamefont
  {Keimer}}, \bibinfo {author} {\bibfnamefont {R.~J.}\ \bibnamefont
  {Birgeneau}}, \bibinfo {author} {\bibfnamefont {A.}~\bibnamefont {Cassanho}},
  \bibinfo {author} {\bibfnamefont {Y.}~\bibnamefont {Endoh}}, \bibinfo
  {author} {\bibfnamefont {M.}~\bibnamefont {Greven}}, \bibinfo {author}
  {\bibfnamefont {M.~A.}\ \bibnamefont {Kastner}},\ and\ \bibinfo {author}
  {\bibfnamefont {G.}~\bibnamefont {Shirane}},\ }\bibfield  {title} {\bibinfo
  {title} {{Soft phonon behavior and magnetism at the low temperature
  structural phase transition of La$_{1.65}$Nd$_{0.35}$CuO$_4$}},\ }\href
  {https://doi.org/10.1007/BF01344067} {\bibfield  {journal} {\bibinfo
  {journal} {Z. Phys. B}\ }\textbf {\bibinfo {volume} {91}},\ \bibinfo {pages}
  {373} (\bibinfo {year} {1993})}\BibitemShut {NoStop}%
\bibitem [{\citenamefont {Yildirim}\ \emph
  {et~al.}(1994{\natexlab{b}})\citenamefont {Yildirim}, \citenamefont {Harris},
  \citenamefont {Entin-Wohlman},\ and\ \citenamefont {Aharony}}]{yild94a}%
  \BibitemOpen
  \bibfield  {author} {\bibinfo {author} {\bibfnamefont {T.}~\bibnamefont
  {Yildirim}}, \bibinfo {author} {\bibfnamefont {A.~B.}\ \bibnamefont
  {Harris}}, \bibinfo {author} {\bibfnamefont {O.}~\bibnamefont
  {Entin-Wohlman}},\ and\ \bibinfo {author} {\bibfnamefont {A.}~\bibnamefont
  {Aharony}},\ }\bibfield  {title} {\bibinfo {title} {{Spin structures of
  tetragonal lamellar copper oxides}},\ }\href
  {https://doi.org/10.1103/PhysRevLett.72.3710} {\bibfield  {journal} {\bibinfo
   {journal} {Phys. Rev. Lett.}\ }\textbf {\bibinfo {volume} {72}},\ \bibinfo
  {pages} {3710} (\bibinfo {year} {1994}{\natexlab{b}})}\BibitemShut {NoStop}%
\bibitem [{\citenamefont {Radaelli}\ \emph {et~al.}(1994)\citenamefont
  {Radaelli}, \citenamefont {Hinks}, \citenamefont {Mitchell}, \citenamefont
  {Hunter}, \citenamefont {Wagner}, \citenamefont {Dabrowski}, \citenamefont
  {Vandervoort}, \citenamefont {Viswanathan},\ and\ \citenamefont
  {Jorgensen}}]{rada94}%
  \BibitemOpen
  \bibfield  {author} {\bibinfo {author} {\bibfnamefont {P.~G.}\ \bibnamefont
  {Radaelli}}, \bibinfo {author} {\bibfnamefont {D.~G.}\ \bibnamefont {Hinks}},
  \bibinfo {author} {\bibfnamefont {A.~W.}\ \bibnamefont {Mitchell}}, \bibinfo
  {author} {\bibfnamefont {B.~A.}\ \bibnamefont {Hunter}}, \bibinfo {author}
  {\bibfnamefont {J.~L.}\ \bibnamefont {Wagner}}, \bibinfo {author}
  {\bibfnamefont {B.}~\bibnamefont {Dabrowski}}, \bibinfo {author}
  {\bibfnamefont {K.~G.}\ \bibnamefont {Vandervoort}}, \bibinfo {author}
  {\bibfnamefont {H.~K.}\ \bibnamefont {Viswanathan}},\ and\ \bibinfo {author}
  {\bibfnamefont {J.~D.}\ \bibnamefont {Jorgensen}},\ }\bibfield  {title}
  {\bibinfo {title} {{Structural and superconducting properties of
  ${\mathrm{La}}_{2\mathrm{-}\mathit{x}}$${\mathrm{Sr}}_{\mathit{x}}$${\mathrm{CuO}}_{4}$
  as a function of Sr content}},\ }\href
  {https://doi.org/10.1103/PhysRevB.49.4163} {\bibfield  {journal} {\bibinfo
  {journal} {Phys. Rev. B}\ }\textbf {\bibinfo {volume} {49}},\ \bibinfo
  {pages} {4163} (\bibinfo {year} {1994})}\BibitemShut {NoStop}%
\bibitem [{\citenamefont {Winn}\ \emph {et~al.}(2015)\citenamefont {Winn},
  \citenamefont {Filges}, \citenamefont {Garlea}, \citenamefont {Graves-Brook},
  \citenamefont {Hagen}, \citenamefont {Jiang}, \citenamefont {Kenzelmann},
  \citenamefont {Passell}, \citenamefont {Shapiro}, \citenamefont {Tong},\ and\
  \citenamefont {Zaliznyak}}]{hyspec15}%
  \BibitemOpen
  \bibfield  {author} {\bibinfo {author} {\bibfnamefont {B.}~\bibnamefont
  {Winn}}, \bibinfo {author} {\bibfnamefont {U.}~\bibnamefont {Filges}},
  \bibinfo {author} {\bibfnamefont {O.~V.}\ \bibnamefont {Garlea}}, \bibinfo
  {author} {\bibfnamefont {M.}~\bibnamefont {Graves-Brook}}, \bibinfo {author}
  {\bibfnamefont {M.}~\bibnamefont {Hagen}}, \bibinfo {author} {\bibfnamefont
  {C.~Y.}\ \bibnamefont {Jiang}}, \bibinfo {author} {\bibfnamefont
  {M.}~\bibnamefont {Kenzelmann}}, \bibinfo {author} {\bibfnamefont
  {L.}~\bibnamefont {Passell}}, \bibinfo {author} {\bibfnamefont {S.~M.}\
  \bibnamefont {Shapiro}}, \bibinfo {author} {\bibfnamefont {X.}~\bibnamefont
  {Tong}},\ and\ \bibinfo {author} {\bibfnamefont {I.}~\bibnamefont
  {Zaliznyak}},\ }\bibfield  {title} {\bibinfo {title} {{Recent progress on
  HYSPEC, and its polarization analysis capabilities}},\ }\href
  {http://dx.doi.org/10.1051/epjconf/20158303017} {\bibfield  {journal}
  {\bibinfo  {journal} {EPJ Web Conf.}\ }\textbf {\bibinfo {volume} {83}},\
  \bibinfo {pages} {03017} (\bibinfo {year} {2015})}\BibitemShut {NoStop}%
\bibitem [{\citenamefont {Walters}\ \emph {et~al.}(2009)\citenamefont
  {Walters}, \citenamefont {Perring}, \citenamefont {Caux}, \citenamefont
  {Savici}, \citenamefont {Gu}, \citenamefont {Lee}, \citenamefont {Ku},\ and\
  \citenamefont {Zaliznyak}}]{walt09}%
  \BibitemOpen
  \bibfield  {author} {\bibinfo {author} {\bibfnamefont {A.~C.}\ \bibnamefont
  {Walters}}, \bibinfo {author} {\bibfnamefont {T.~G.}\ \bibnamefont
  {Perring}}, \bibinfo {author} {\bibfnamefont {J.-S.}\ \bibnamefont {Caux}},
  \bibinfo {author} {\bibfnamefont {A.~T.}\ \bibnamefont {Savici}}, \bibinfo
  {author} {\bibfnamefont {G.~D.}\ \bibnamefont {Gu}}, \bibinfo {author}
  {\bibfnamefont {C.-C.}\ \bibnamefont {Lee}}, \bibinfo {author} {\bibfnamefont
  {W.}~\bibnamefont {Ku}},\ and\ \bibinfo {author} {\bibfnamefont {I.~A.}\
  \bibnamefont {Zaliznyak}},\ }\bibfield  {title} {\bibinfo {title} {{Effect of
  covalent bonding on magnetism and the missing neutron intensity in copper
  oxide compounds}},\ }\href {https://doi.org/10.1038/nphys1405} {\bibfield
  {journal} {\bibinfo  {journal} {Nat. Phys.}\ }\textbf {\bibinfo {volume}
  {5}},\ \bibinfo {pages} {867} (\bibinfo {year} {2009})}\BibitemShut {NoStop}%
\bibitem [{Note1()}]{Note1}%
  \BibitemOpen
  \bibinfo {note} {For the magnetic form factor, we use $f_{\protect \rm
  Cu}(1,0,0) = f_{\protect \rm Cu}(0,1,0) = 0.63$, $f_{\protect \rm Cu}(2,1,0)
  = f_{\protect \rm Cu}(1,2,0) = 0.55$, and $f_{\protect \rm Cu}(3,0,0) =
  0.50$.}\BibitemShut {Stop}%
\bibitem [{\citenamefont {Silva~Neto}\ \emph {et~al.}(2006)\citenamefont
  {Silva~Neto}, \citenamefont {Benfatto}, \citenamefont {Juricic},\ and\
  \citenamefont {Morais~Smith}}]{silv06}%
  \BibitemOpen
  \bibfield  {author} {\bibinfo {author} {\bibfnamefont {M.~B.}\ \bibnamefont
  {Silva~Neto}}, \bibinfo {author} {\bibfnamefont {L.}~\bibnamefont
  {Benfatto}}, \bibinfo {author} {\bibfnamefont {V.}~\bibnamefont {Juricic}},\
  and\ \bibinfo {author} {\bibfnamefont {C.}~\bibnamefont {Morais~Smith}},\
  }\bibfield  {title} {\bibinfo {title} {{Magnetic susceptibility anisotropies
  in a two-dimensional quantum Heisenberg antiferromagnet with
  Dzyaloshinskii-Moriya interactions}},\ }\href
  {https://doi.org/10.1103/PhysRevB.73.045132} {\bibfield  {journal} {\bibinfo
  {journal} {Phys. Rev. B}\ }\textbf {\bibinfo {volume} {73}},\ \bibinfo
  {pages} {045132} (\bibinfo {year} {2006})}\BibitemShut {NoStop}%
\bibitem [{\citenamefont {Fujita}\ \emph {et~al.}(2004)\citenamefont {Fujita},
  \citenamefont {Goka}, \citenamefont {Yamada}, \citenamefont {Tranquada},\
  and\ \citenamefont {Regnault}}]{fuji04}%
  \BibitemOpen
  \bibfield  {author} {\bibinfo {author} {\bibfnamefont {M.}~\bibnamefont
  {Fujita}}, \bibinfo {author} {\bibfnamefont {H.}~\bibnamefont {Goka}},
  \bibinfo {author} {\bibfnamefont {K.}~\bibnamefont {Yamada}}, \bibinfo
  {author} {\bibfnamefont {J.~M.}\ \bibnamefont {Tranquada}},\ and\ \bibinfo
  {author} {\bibfnamefont {L.~P.}\ \bibnamefont {Regnault}},\ }\bibfield
  {title} {\bibinfo {title} {{Stripe order, depinning, and fluctuations in
  ${\mathrm{La}}_{1.875}{\mathrm{Ba}}_{0.125}{\mathrm{CuO}}_{4}$ and
  ${\mathrm{La}}_{1.875}{\mathrm{Ba}}_{0.075}{\mathrm{Sr}}_{0.050}{\mathrm{CuO}}_{4}$}},\
  }\href {https://doi.org/10.1103/PhysRevB.70.104517} {\bibfield  {journal}
  {\bibinfo  {journal} {Phys. Rev. B}\ }\textbf {\bibinfo {volume} {70}},\
  \bibinfo {pages} {104517} (\bibinfo {year} {2004})}\BibitemShut {NoStop}%
\bibitem [{\citenamefont {Axe}\ \emph {et~al.}(1989)\citenamefont {Axe},
  \citenamefont {Moudden}, \citenamefont {Hohlwein}, \citenamefont {Cox},
  \citenamefont {Mohanty}, \citenamefont {Moodenbaugh},\ and\ \citenamefont
  {Xu}}]{axe89}%
  \BibitemOpen
  \bibfield  {author} {\bibinfo {author} {\bibfnamefont {J.~D.}\ \bibnamefont
  {Axe}}, \bibinfo {author} {\bibfnamefont {A.~H.}\ \bibnamefont {Moudden}},
  \bibinfo {author} {\bibfnamefont {D.}~\bibnamefont {Hohlwein}}, \bibinfo
  {author} {\bibfnamefont {D.~E.}\ \bibnamefont {Cox}}, \bibinfo {author}
  {\bibfnamefont {K.~M.}\ \bibnamefont {Mohanty}}, \bibinfo {author}
  {\bibfnamefont {A.~R.}\ \bibnamefont {Moodenbaugh}},\ and\ \bibinfo {author}
  {\bibfnamefont {Y.}~\bibnamefont {Xu}},\ }\bibfield  {title} {\bibinfo
  {title} {{Structural phase transformations and superconductivity in
  ${\mathrm{La}}_{2\mathrm{\ensuremath{-}}\mathrm{x}}$${\mathrm{Ba}}_{\mathrm{x}}$${\mathrm{CuO}}_{4}$}},\
  }\href {https://doi.org/10.1103/PhysRevLett.62.2751} {\bibfield  {journal}
  {\bibinfo  {journal} {Phys. Rev. Lett.}\ }\textbf {\bibinfo {volume} {62}},\
  \bibinfo {pages} {2751} (\bibinfo {year} {1989})}\BibitemShut {NoStop}%
\bibitem [{\citenamefont {H\"ucker}\ \emph {et~al.}(2011)\citenamefont
  {H\"ucker}, \citenamefont {v.~Zimmermann}, \citenamefont {Gu}, \citenamefont
  {Xu}, \citenamefont {Wen}, \citenamefont {Xu}, \citenamefont {Kang},
  \citenamefont {Zheludev},\ and\ \citenamefont {Tranquada}}]{huck11}%
  \BibitemOpen
  \bibfield  {author} {\bibinfo {author} {\bibfnamefont {M.}~\bibnamefont
  {H\"ucker}}, \bibinfo {author} {\bibfnamefont {M.}~\bibnamefont
  {v.~Zimmermann}}, \bibinfo {author} {\bibfnamefont {G.~D.}\ \bibnamefont
  {Gu}}, \bibinfo {author} {\bibfnamefont {Z.~J.}\ \bibnamefont {Xu}}, \bibinfo
  {author} {\bibfnamefont {J.~S.}\ \bibnamefont {Wen}}, \bibinfo {author}
  {\bibfnamefont {G.}~\bibnamefont {Xu}}, \bibinfo {author} {\bibfnamefont
  {H.~J.}\ \bibnamefont {Kang}}, \bibinfo {author} {\bibfnamefont
  {A.}~\bibnamefont {Zheludev}},\ and\ \bibinfo {author} {\bibfnamefont
  {J.~M.}\ \bibnamefont {Tranquada}},\ }\bibfield  {title} {\bibinfo {title}
  {{Stripe order in superconducting La$_{2-x}$Ba$_{x}$CuO$_{4}$
  ($0.095\le{}x\le{}0.155$)}},\ }\href
  {https://doi.org/10.1103/PhysRevB.83.104506} {\bibfield  {journal} {\bibinfo
  {journal} {Phys. Rev. B}\ }\textbf {\bibinfo {volume} {83}},\ \bibinfo
  {pages} {104506} (\bibinfo {year} {2011})}\BibitemShut {NoStop}%
\bibitem [{\citenamefont {Tworzyd\l{}o}\ \emph {et~al.}(1999)\citenamefont
  {Tworzyd\l{}o}, \citenamefont {Osman}, \citenamefont {van Duin},\ and\
  \citenamefont {Zaanen}}]{twor99}%
  \BibitemOpen
  \bibfield  {author} {\bibinfo {author} {\bibfnamefont {J.}~\bibnamefont
  {Tworzyd\l{}o}}, \bibinfo {author} {\bibfnamefont {O.~Y.}\ \bibnamefont
  {Osman}}, \bibinfo {author} {\bibfnamefont {C.~N.~A.}\ \bibnamefont {van
  Duin}},\ and\ \bibinfo {author} {\bibfnamefont {J.}~\bibnamefont {Zaanen}},\
  }\bibfield  {title} {\bibinfo {title} {{Quantum magnetism in the stripe
  phase: Bond versus site order}},\ }\href
  {https://doi.org/10.1103/PhysRevB.59.115} {\bibfield  {journal} {\bibinfo
  {journal} {Phys. Rev. B}\ }\textbf {\bibinfo {volume} {59}},\ \bibinfo
  {pages} {115} (\bibinfo {year} {1999})}\BibitemShut {NoStop}%
\bibitem [{\citenamefont {Yao}\ \emph {et~al.}(2006)\citenamefont {Yao},
  \citenamefont {Carlson},\ and\ \citenamefont {Campbell}}]{yao06a}%
  \BibitemOpen
  \bibfield  {author} {\bibinfo {author} {\bibfnamefont {D.~X.}\ \bibnamefont
  {Yao}}, \bibinfo {author} {\bibfnamefont {E.~W.}\ \bibnamefont {Carlson}},\
  and\ \bibinfo {author} {\bibfnamefont {D.~K.}\ \bibnamefont {Campbell}},\
  }\bibfield  {title} {\bibinfo {title} {{Magnetic Excitations of Stripes near
  a Quantum Critical Point}},\ }\href
  {https://doi.org/10.1103/PhysRevLett.97.017003} {\bibfield  {journal}
  {\bibinfo  {journal} {Phys. Rev. Lett.}\ }\textbf {\bibinfo {volume} {97}},\
  \bibinfo {pages} {017003} (\bibinfo {year} {2006})}\BibitemShut {NoStop}%
\bibitem [{\citenamefont {Tranquada}\ \emph {et~al.}(2008)\citenamefont
  {Tranquada}, \citenamefont {Gu}, \citenamefont {H{\"u}cker}, \citenamefont
  {Jie}, \citenamefont {Kang}, \citenamefont {Klingeler}, \citenamefont {Li},
  \citenamefont {Tristan}, \citenamefont {Wen}, \citenamefont {Xu},
  \citenamefont {Xu}, \citenamefont {Zhou},\ and\ \citenamefont
  {v.~Zimmermann}}]{tran08}%
  \BibitemOpen
  \bibfield  {author} {\bibinfo {author} {\bibfnamefont {J.~M.}\ \bibnamefont
  {Tranquada}}, \bibinfo {author} {\bibfnamefont {G.~D.}\ \bibnamefont {Gu}},
  \bibinfo {author} {\bibfnamefont {M.}~\bibnamefont {H{\"u}cker}}, \bibinfo
  {author} {\bibfnamefont {Q.}~\bibnamefont {Jie}}, \bibinfo {author}
  {\bibfnamefont {H.-J.}\ \bibnamefont {Kang}}, \bibinfo {author}
  {\bibfnamefont {R.}~\bibnamefont {Klingeler}}, \bibinfo {author}
  {\bibfnamefont {Q.}~\bibnamefont {Li}}, \bibinfo {author} {\bibfnamefont
  {N.}~\bibnamefont {Tristan}}, \bibinfo {author} {\bibfnamefont {J.~S.}\
  \bibnamefont {Wen}}, \bibinfo {author} {\bibfnamefont {G.~Y.}\ \bibnamefont
  {Xu}}, \bibinfo {author} {\bibfnamefont {Z.~J.}\ \bibnamefont {Xu}}, \bibinfo
  {author} {\bibfnamefont {J.}~\bibnamefont {Zhou}},\ and\ \bibinfo {author}
  {\bibfnamefont {M.}~\bibnamefont {v.~Zimmermann}},\ }\bibfield  {title}
  {\bibinfo {title} {{Evidence for unusual superconducting correlations
  coexisting with stripe order in La$_{1.875}$Ba$_{0.125}$CuO$_4$}},\
  }\href@noop {} {\bibfield  {journal} {\bibinfo  {journal} {Phys. Rev. B}\
  }\textbf {\bibinfo {volume} {78}},\ \bibinfo {eid} {174529} (\bibinfo {year}
  {2008})}\BibitemShut {NoStop}%
\bibitem [{\citenamefont {H\"ucker}\ \emph {et~al.}(2008)\citenamefont
  {H\"ucker}, \citenamefont {Gu},\ and\ \citenamefont {Tranquada}}]{huck08}%
  \BibitemOpen
  \bibfield  {author} {\bibinfo {author} {\bibfnamefont {M.}~\bibnamefont
  {H\"ucker}}, \bibinfo {author} {\bibfnamefont {G.~D.}\ \bibnamefont {Gu}},\
  and\ \bibinfo {author} {\bibfnamefont {J.~M.}\ \bibnamefont {Tranquada}},\
  }\bibfield  {title} {\bibinfo {title} {{Spin susceptibility of underdoped
  cuprate superconductors: Insights from a stripe-ordered crystal}},\ }\href
  {https://doi.org/10.1103/PhysRevB.78.214507} {\bibfield  {journal} {\bibinfo
  {journal} {Phys. Rev. B}\ }\textbf {\bibinfo {volume} {78}},\ \bibinfo
  {pages} {214507} (\bibinfo {year} {2008})}\BibitemShut {NoStop}%
\bibitem [{\citenamefont {Karapetyan}\ \emph {et~al.}(2012)\citenamefont
  {Karapetyan}, \citenamefont {H\"ucker}, \citenamefont {Gu}, \citenamefont
  {Tranquada}, \citenamefont {Fejer}, \citenamefont {Xia},\ and\ \citenamefont
  {Kapitulnik}}]{kara12}%
  \BibitemOpen
  \bibfield  {author} {\bibinfo {author} {\bibfnamefont {H.}~\bibnamefont
  {Karapetyan}}, \bibinfo {author} {\bibfnamefont {M.}~\bibnamefont
  {H\"ucker}}, \bibinfo {author} {\bibfnamefont {G.~D.}\ \bibnamefont {Gu}},
  \bibinfo {author} {\bibfnamefont {J.~M.}\ \bibnamefont {Tranquada}}, \bibinfo
  {author} {\bibfnamefont {M.~M.}\ \bibnamefont {Fejer}}, \bibinfo {author}
  {\bibfnamefont {J.}~\bibnamefont {Xia}},\ and\ \bibinfo {author}
  {\bibfnamefont {A.}~\bibnamefont {Kapitulnik}},\ }\bibfield  {title}
  {\bibinfo {title} {{Magneto-Optical Measurements of a Cascade of Transitions
  in Superconducting
  ${\mathrm{La}}_{1.875}{\mathrm{Ba}}_{0.125}{\mathrm{CuO}}_{4}$ Single
  Crystals}},\ }\href@noop {} {\bibfield  {journal} {\bibinfo  {journal} {Phys.
  Rev. Lett.}\ }\textbf {\bibinfo {volume} {109}},\ \bibinfo {pages} {147001}
  (\bibinfo {year} {2012})}\BibitemShut {NoStop}%
\bibitem [{\citenamefont {Wilkins}\ \emph {et~al.}(2011)\citenamefont
  {Wilkins}, \citenamefont {Dean}, \citenamefont {Fink}, \citenamefont
  {H\"ucker}, \citenamefont {Geck}, \citenamefont {Soltwisch}, \citenamefont
  {Schierle}, \citenamefont {Weschke}, \citenamefont {Gu}, \citenamefont
  {Uchida}, \citenamefont {Ichikawa}, \citenamefont {Tranquada},\ and\
  \citenamefont {Hill}}]{wilk11}%
  \BibitemOpen
  \bibfield  {author} {\bibinfo {author} {\bibfnamefont {S.~B.}\ \bibnamefont
  {Wilkins}}, \bibinfo {author} {\bibfnamefont {M.~P.~M.}\ \bibnamefont
  {Dean}}, \bibinfo {author} {\bibfnamefont {J.}~\bibnamefont {Fink}}, \bibinfo
  {author} {\bibfnamefont {M.}~\bibnamefont {H\"ucker}}, \bibinfo {author}
  {\bibfnamefont {J.}~\bibnamefont {Geck}}, \bibinfo {author} {\bibfnamefont
  {V.}~\bibnamefont {Soltwisch}}, \bibinfo {author} {\bibfnamefont
  {E.}~\bibnamefont {Schierle}}, \bibinfo {author} {\bibfnamefont
  {E.}~\bibnamefont {Weschke}}, \bibinfo {author} {\bibfnamefont
  {G.}~\bibnamefont {Gu}}, \bibinfo {author} {\bibfnamefont {S.}~\bibnamefont
  {Uchida}}, \bibinfo {author} {\bibfnamefont {N.}~\bibnamefont {Ichikawa}},
  \bibinfo {author} {\bibfnamefont {J.~M.}\ \bibnamefont {Tranquada}},\ and\
  \bibinfo {author} {\bibfnamefont {J.~P.}\ \bibnamefont {Hill}},\ }\bibfield
  {title} {\bibinfo {title} {{Comparison of stripe modulations in
  La$_{1.875}$Ba$_{0.125}$CuO$_{4}$ and
  La$_{1.48}$Nd$_{0.4}$Sr$_{0.12}$CuO$_{4}$}},\ }\href@noop {} {\bibfield
  {journal} {\bibinfo  {journal} {Phys. Rev. B}\ }\textbf {\bibinfo {volume}
  {84}},\ \bibinfo {pages} {195101} (\bibinfo {year} {2011})}\BibitemShut
  {NoStop}%
\bibitem [{\citenamefont {Orenstein}(2011)}]{oren11}%
  \BibitemOpen
  \bibfield  {author} {\bibinfo {author} {\bibfnamefont {J.}~\bibnamefont
  {Orenstein}},\ }\bibfield  {title} {\bibinfo {title} {{Optical Nonreciprocity
  in Magnetic Structures Related to High-${T}_{c}$ Superconductors}},\ }\href
  {https://doi.org/10.1103/PhysRevLett.107.067002} {\bibfield  {journal}
  {\bibinfo  {journal} {Phys. Rev. Lett.}\ }\textbf {\bibinfo {volume} {107}},\
  \bibinfo {pages} {067002} (\bibinfo {year} {2011})}\BibitemShut {NoStop}%
\bibitem [{\citenamefont {Fried}(2014)}]{frie14}%
  \BibitemOpen
  \bibfield  {author} {\bibinfo {author} {\bibfnamefont {A.~D.}\ \bibnamefont
  {Fried}},\ }\bibfield  {title} {\bibinfo {title} {{Relationship of
  time-reversal symmetry breaking to optical Kerr rotation}},\ }\href
  {https://doi.org/10.1103/PhysRevB.90.121112} {\bibfield  {journal} {\bibinfo
  {journal} {Phys. Rev. B}\ }\textbf {\bibinfo {volume} {90}},\ \bibinfo
  {pages} {121112(R)} (\bibinfo {year} {2014})}\BibitemShut {NoStop}%
\bibitem [{\citenamefont {Karapetyan}\ \emph {et~al.}(2014)\citenamefont
  {Karapetyan}, \citenamefont {Xia}, \citenamefont {H\"ucker}, \citenamefont
  {Gu}, \citenamefont {Tranquada}, \citenamefont {Fejer},\ and\ \citenamefont
  {Kapitulnik}}]{kara14}%
  \BibitemOpen
  \bibfield  {author} {\bibinfo {author} {\bibfnamefont {H.}~\bibnamefont
  {Karapetyan}}, \bibinfo {author} {\bibfnamefont {J.}~\bibnamefont {Xia}},
  \bibinfo {author} {\bibfnamefont {M.}~\bibnamefont {H\"ucker}}, \bibinfo
  {author} {\bibfnamefont {G.~D.}\ \bibnamefont {Gu}}, \bibinfo {author}
  {\bibfnamefont {J.~M.}\ \bibnamefont {Tranquada}}, \bibinfo {author}
  {\bibfnamefont {M.~M.}\ \bibnamefont {Fejer}},\ and\ \bibinfo {author}
  {\bibfnamefont {A.}~\bibnamefont {Kapitulnik}},\ }\bibfield  {title}
  {\bibinfo {title} {{Evidence of Chiral Order in the Charge-Ordered Phase of
  Superconducting
  ${\mathrm{La}}_{1.875}{\mathrm{Ba}}_{0.125}{\mathrm{Cuo}}_{4}$ Single
  Crystals Using Polar Kerr-Effect Measurements}},\ }\href
  {https://doi.org/10.1103/PhysRevLett.112.047003} {\bibfield  {journal}
  {\bibinfo  {journal} {Phys. Rev. Lett.}\ }\textbf {\bibinfo {volume} {112}},\
  \bibinfo {pages} {047003} (\bibinfo {year} {2014})}\BibitemShut {NoStop}%
\bibitem [{\citenamefont {Wang}\ \emph {et~al.}(2014)\citenamefont {Wang},
  \citenamefont {Chubukov},\ and\ \citenamefont {Nandkishore}}]{wang14}%
  \BibitemOpen
  \bibfield  {author} {\bibinfo {author} {\bibfnamefont {Y.}~\bibnamefont
  {Wang}}, \bibinfo {author} {\bibfnamefont {A.}~\bibnamefont {Chubukov}},\
  and\ \bibinfo {author} {\bibfnamefont {R.}~\bibnamefont {Nandkishore}},\
  }\bibfield  {title} {\bibinfo {title} {{Polar Kerr effect from chiral-nematic
  charge order}},\ }\href {https://doi.org/10.1103/PhysRevB.90.205130}
  {\bibfield  {journal} {\bibinfo  {journal} {Phys. Rev. B}\ }\textbf {\bibinfo
  {volume} {90}},\ \bibinfo {pages} {205130} (\bibinfo {year}
  {2014})}\BibitemShut {NoStop}%
\bibitem [{\citenamefont {Orenstein}\ and\ \citenamefont
  {Moore}(2013)}]{oren13}%
  \BibitemOpen
  \bibfield  {author} {\bibinfo {author} {\bibfnamefont {J.}~\bibnamefont
  {Orenstein}}\ and\ \bibinfo {author} {\bibfnamefont {J.~E.}\ \bibnamefont
  {Moore}},\ }\bibfield  {title} {\bibinfo {title} {{Berry phase mechanism for
  optical gyrotropy in stripe-ordered cuprates}},\ }\href
  {https://doi.org/10.1103/PhysRevB.87.165110} {\bibfield  {journal} {\bibinfo
  {journal} {Phys. Rev. B}\ }\textbf {\bibinfo {volume} {87}},\ \bibinfo
  {pages} {165110} (\bibinfo {year} {2013})}\BibitemShut {NoStop}%
\bibitem [{\citenamefont {Sears}\ \emph {et~al.}(2023)\citenamefont {Sears},
  \citenamefont {Shen}, \citenamefont {Krogstad}, \citenamefont {Miao},
  \citenamefont {Bozin}, \citenamefont {Robinson}, \citenamefont {Gu},
  \citenamefont {Osborn}, \citenamefont {Rosenkranz}, \citenamefont
  {Tranquada},\ and\ \citenamefont {Dean}}]{sear23}%
  \BibitemOpen
  \bibfield  {author} {\bibinfo {author} {\bibfnamefont {J.}~\bibnamefont
  {Sears}}, \bibinfo {author} {\bibfnamefont {Y.}~\bibnamefont {Shen}},
  \bibinfo {author} {\bibfnamefont {M.~J.}\ \bibnamefont {Krogstad}}, \bibinfo
  {author} {\bibfnamefont {H.}~\bibnamefont {Miao}}, \bibinfo {author}
  {\bibfnamefont {E.~S.}\ \bibnamefont {Bozin}}, \bibinfo {author}
  {\bibfnamefont {I.~K.}\ \bibnamefont {Robinson}}, \bibinfo {author}
  {\bibfnamefont {G.~D.}\ \bibnamefont {Gu}}, \bibinfo {author} {\bibfnamefont
  {R.}~\bibnamefont {Osborn}}, \bibinfo {author} {\bibfnamefont
  {S.}~\bibnamefont {Rosenkranz}}, \bibinfo {author} {\bibfnamefont {J.~M.}\
  \bibnamefont {Tranquada}},\ and\ \bibinfo {author} {\bibfnamefont {M.~P.~M.}\
  \bibnamefont {Dean}},\ }\bibfield  {title} {\bibinfo {title} {{Structure of
  charge density waves in
  ${\mathrm{La}}_{1.875}{\mathrm{Ba}}_{0.125}{\mathrm{CuO}}_{4}$}},\ }\href
  {https://doi.org/10.1103/PhysRevB.107.115125} {\bibfield  {journal} {\bibinfo
   {journal} {Phys. Rev. B}\ }\textbf {\bibinfo {volume} {107}},\ \bibinfo
  {pages} {115125} (\bibinfo {year} {2023})}\BibitemShut {NoStop}%
\end{thebibliography}%

\end{document}